\documentclass[10pt]{article}

\usepackage[margin=0.85in]{geometry}
\usepackage{amsmath, amssymb}
\usepackage{natbib}
\usepackage{booktabs}
\usepackage{graphicx}
\usepackage{xcolor}
\usepackage{titling}
\usepackage{placeins}
\usepackage[hidelinks]{hyperref}

\title{Mitigating Emergent Collusion in LLM Pricing Agents}
\author{
Abdullah Garra \thanks{This work was conducted at Tel Aviv University.}\\
University of Massachusetts Amherst\\
\texttt{agarrah@umass.edu}
}
\date{}

\begin{document}

\maketitle

\begin{abstract}
Recent work shows that LLM-based pricing agents can produce supracompetitive outcomes in repeated oligopoly environments without being explicitly instructed to collude. We reproduce the qualitative prompt-sensitivity effect of Fish et al. using DeepSeek-V3.1: the P1 prompt produces significantly higher prices and profits than P2, although our outcomes are less monopoly-like than the original GPT-4 results. We then evaluate three regulatory interventions: a prompt-only warning, a Harrington-inspired expected-damages payoff regulator, and an active random entrant. The prompt-only regulator reduces but does not eliminate above-Nash pricing. The Harrington regulator brings P1 outcomes close to the duopoly Nash benchmark and removes the statistically significant P1--P2 gap. The active entrant produces the strongest effect, pushing both prompts below the appropriate random-entrant Nash benchmark. Overall, our experiments provide preliminary evidence that interventions that alter incentives or market participation can reduce supracompetitive pricing more effectively than prompt warnings alone.
\end{abstract}

\section{Algorithmic Pricing Collusion in LLMs}

\citet{fish2025} study whether Large Language Model (LLM)-based pricing agents can autonomously generate collusive-like outcomes in repeated market environments. Their central object of study is \emph{autonomous algorithmic collusion}: supracompetitive outcomes produced by learning or AI-based algorithms without explicit human design to collude. This framing builds on a growing literature on algorithmic pricing and competition policy, where regulators and researchers have raised concerns that automated pricing systems may sustain high prices even without direct communication among firms \citep{harrington2018, ezrachi2020, calvano2020b, abada2024, brown2021}.

The paper situates itself against earlier work on algorithmic collusion, especially studies of Q-learning pricing agents. \citet{calvano2020b} show that Q-learning algorithms can learn supracompetitive pricing in repeated Bertrand environments. Fish et al. emphasize, however, that these earlier algorithmic agents face practical barriers where they often require long training periods, and may be exploitable by competitors. This motivates the paper's key policy question which can be summed as: do modern AI systems that avoid these barriers also exhibit autonomous algorithmic collusion?

The authors argue that LLMs are an especially important test case. Unlike classical reinforcement-learning pricing algorithms, LLMs are pretrained on broad datasets and can be deployed without costly environment-specific training. They can also operate in many different environments using natural-language instructions. These features make them plausible candidates for real-world pricing agents. At the same time, however, they are opaque, stochastic, and difficult to interpret, which creates both scientific and regulatory challenges. The paper therefore treats LLMs not merely as tools for simulating humans, but as a new kind of economic agent whose behavior must be experimentally analyzed.

Fish et al. explicitly ask three research questions: whether LLM-based pricing agents behave in a collusive-like manner and reach supracompetitive prices; whether they do so without being instructed or intended to collude; and, if so, what mechanisms drive this behavior. They claim to be the first to experimentally study autonomous algorithmic collusion by LLM-based pricing agents.

\subsection{Experimental Design}
\label{sec:fish_experiment}
The main experiment uses a repeated Bertrand oligopoly environment following \citet{calvano2020b}. Each firm delegates pricing to an independent LLM-based agent. In every period, each agent sets a price, demand is realized, and the agent observes the prices chosen by all firms as well as its own quantity sold and profit. Agents do not observe the underlying demand function and cannot communicate directly or indirectly except through the prices they set.

The environment uses a logit demand model. If firms $1,\ldots,n$ set prices $p_1,\ldots,p_n$, firm $i$'s demand is
\[
q_i =
\beta
\frac{
\exp((a_i-p_i/\alpha)/\mu)
}{
\sum_{j=1}^n \exp((a_j-p_j/\alpha)/\mu) + \exp(a_0/\mu)
}.
\]
Profits are given by
\[
\pi_i = (p_i-\alpha c_i)q_i.
\]
In the main experiments, the authors use symmetric firms with $a_i=2$, outside option $a_0=0$, horizontal differentiation parameter $\mu=0.25$, and marginal cost $c_i=1$, following \citet{calvano2020b}. They set $\beta=100$ so that quantities appear in a natural numerical scale to the LLM. They also randomize the currency-scaling parameter $\alpha \in \{1,3.2,10\}$ because LLMs may not be invariant to changes in units.

A single experimental run consists of 300 periods and the LLM prompt contains five main components: a prompt prefix stating the high-level goal, basic market information including marginal cost and a price ceiling, market history from the previous 100 periods, the agent's own ``plans and insights'' from the previous period, and output instructions requiring the agent to generate reasoning, update its plans, and set a price. This design gives the otherwise memoryless LLM some continuity of thought across periods.

The authors compare two prompt prefixes, denoted P1 and P2. Both instruct the agent to maximize long-run profit, but they differ in their final phrasing. P1 emphasizes avoiding actions that undermine profitability, while P2 explicitly mentions exploration, risky or aggressive pricing, and the fact that pricing lower than a competitor can increase quantity sold. The exact prompt summaries are given in Appendix~\ref{app:prompts}. This prompt manipulation is important because one of the paper's claims is that seemingly innocuous language can substantially alter market outcomes.

Before running the main duopoly experiment, the authors evaluate candidate LLMs in a monopoly setting. They view strong monopoly performance as a necessary condition for meaningful multi-firm pricing experiments. Based on this preliminary evaluation, they select GPT-4 as the primary model for the main duopoly experiment.

For each prompt prefix, P1 and P2, the authors conduct 21 independent 300-period duopoly runs. They measure pricing and profits primarily over the final 50 periods, comparing outcomes to the Bertrand--Nash price $p^{Nash}$ of the one-shot game and to the monopoly price $p^M$ that would maximize total profit if both firms were controlled by a single monopolist. Refer to Appendix \ref{app:benchmarks} for more details on how we compute those in our specific setup.

\subsection{Findings}
The central finding is that LLM-based pricing agents quickly and consistently reach supracompetitive prices and profits. Both P1 and P2 lead to prices above the Bertrand--Nash benchmark. However, P1 produces substantially higher prices than P2, sometimes even above the monopoly price. Profits follow the same pattern: both prompts generate supracompetitive profits, but P1 produces total profits closer to monopoly levels. Notably, neither prompt instructs the agents to collude, retaliate, avoid competition, or coordinate with competitors.

The contrast between P1 and P2 is one of the paper's most important empirical observations. P2 includes language that makes undercutting more salient by telling the agent that lower prices can increase quantity sold. P1 instead reinforces long-run profitability. The authors argue that the difference is not simply that P2 gives the agent new economic knowledge and they show that P1 agents already understand that lower prices can increase quantity sold. Rather, the prompt wording appears to change the strategic frame through which agents interpret the environment.

The authors also report robustness results in appendices. Their main findings persist under noise, asymmetric demand, heterogeneous pricing agents, instructions to discount the future, and newer-generation LLMs. They also extend the analysis to first-price auctions, another domain where prior work found autonomous collusion among Q-learning agents~\citep{banchio2022}.

\subsection{Reward-Punishment and Textual Reasoning}
\label{sec:fish_reward_punishment}
After documenting supracompetitive pricing, Fish et al. ask what mechanisms sustain it. They focus on reward-punishment strategies, a classic explanation for tacit collusion in repeated oligopoly games. In such strategies, agents maintain high prices when rivals cooperate but punish deviations, such as price cuts, with lower prices for some time. This logic appears in the theoretical literature on collusion \citep{stigler1964, friedman1971, green1984, harrington2018} and in prior work on Q-learning pricing agents \citep{calvano2020b}.

The authors conduct two complementary analyses tailored for the LLM case. First, they conduct an off-path analysis that uses the fact that LLM agents produce textual reasoning and plans. The authors search these generated plans for price-war-related reasoning. They find that agents, especially under P1, more often express concern about avoiding price wars. However, textual correlation alone does not establish that such reasoning causally affects pricing.

To address this, the authors introduce an ``implantation'' method. They reset completed simulations to earlier states and replace an agent's plans and insights with text expressing concern about avoiding a price war. They then rerun the experiment from that point and compare the resulting price to the original price. Prices after implantation are significantly higher, which suggests that price-war-concerned reasoning causally contributes to higher pricing. This method is one of the paper's methodological contributions: because AI agents can be reset and perturbed, researchers can perform counterfactual behavioral experiments that are impossible with human subjects.

\paragraph{On-path analysis.} Second, they use regression to examine whether observed prices follow a reward-punishment pattern. The authors estimate a model of the form
\begin{equation}
\label{eq:price_dynamics}
p^t_{i,r} = \alpha_{i,r} + \gamma p^{t-1}_{i,r} + \delta p^{t-1}_{-i,r} + \varepsilon^t_{i,r}
\end{equation}
where $p^t_{i,r}$ is agent $i$'s price in period $t$ of run $r$, $p^{t-1}_{i,r}$ is its own previous price, $p^{t-1}_{-i,r}$ is its competitor's previous price, and $\alpha_{i,r}$ is a firm-run fixed effect. A positive coefficient on the competitor's previous price indicates that the agent raises prices after the competitor raises prices and lowers prices after the competitor lowers prices. A positive coefficient on the agent's own previous price indicates persistence, meaning rewards and punishments decay gradually rather than disappearing immediately.

The regression results are consistent with reward-punishment behavior. Agents respond positively to competitors' previous prices, and their own prices are sticky over time. These effects are stronger under P1, the prompt that also produces higher prices, higher profits, and greater price-war concern in the textual analysis. Taken together, the off-path and on-path evidence suggests that LLM-based pricing agents avoid price reductions partly because they anticipate or ``fear'' retaliation, and when lower prices occur, their realized behavior resembles punishment dynamics.

Finally, the authors broaden the textual analysis using clustering methods. They find that prompt wording systematically changes the types of reasoning agents produce. P1 agents more often emphasize sustaining price levels and reacting to competitors, while P2 agents more often emphasize undercutting and exploration. Using the implantation method again, the authors link undercutting-oriented reasoning to lower prices. Thus, the paper argues not only that LLM agents produce supracompetitive outcomes, but also that their natural-language reasoning contains strategically meaningful content that can help explain their behavior.

Overall, Fish et al. provide evidence that LLM-based pricing agents can autonomously generate collusive-like outcomes in repeated oligopoly environments. Their contribution is not only empirical but also methodological. Empirically, they show that LLM agents reach supracompetitive prices and profits without being instructed to collude, and that small prompt differences can substantially change the degree of supracompetitive pricing. They also show how LLM-generated reasoning can be analyzed and causally perturbed through implantation experiments. The paper therefore raises a regulatory concern: if natural-language prompts can meaningfully affect market conduct, then regulating LLM-based pricing agents may require attention not only to outcomes and code, but also to instructions, reasoning traces, and behavioral responses across market histories.

\section{Study Overview}

We take Fish et al.'s regulatory concern as the starting point for our study. If LLM-based pricing agents can produce supracompetitive outcomes through natural-language instructions and market histories, then a natural next question is: 

\begin{center}
    \textit{Can regulatory interventions reduce prompt-induced supracompetitive pricing by LLM pricing agents, and which intervention channels perform best?}
\end{center}

Thus, our work has two goals. First, we attempt to reproduce the qualitative P1/P2 effect reported by Fish et al. using a lower-cost model due to computational-budget constraints. Second, after establishing a baseline, we introduce several regulatory interventions and evaluate whether they reduce supracompetitive pricing.

The interventions we study target three regulatory channels: modifying the agents' instructions, modifying realized payoffs through expected damages, and changing the market structure through an economically active additional participant.

The remainder of the paper is organized as follows. In Section \ref{sec:related_work}, we situate the study within the literatures on algorithmic collusion, LLM agents, and regulation of pricing algorithms. In Section \ref{sec:reproduction}, we describe our model-selection constraints and try to reproduce the original P1/P2 experiment and results. In Section \ref{sec:regulations}, we define the regulatory interventions we test. Finally, in Section \ref{sec:results}, we report the results, the effects of the interventions, and trajectory-level analyses inspired by Fish et al.'s methodology.

\section{Related Work}
\label{sec:related_work}

Our work builds on three lines of research.

\paragraph{Algorithmic pricing and LLM collusion.}
\citet{calvano2020b} provide the canonical simulation evidence that
Q-learning pricing agents can sustain supracompetitive prices in repeated oligopoly
environments without explicit communication. \citet{fish2025}
extend this concern to LLM-based pricing agents, using an economic environment that
closely follows Calvano et al. but replacing task-specific reinforcement learning with
pretrained, prompt-driven agents that use natural-language plans and memory. Related work also studies sequential pricing and platform-level interventions for algorithmic sellers~\citep{klein2021,johnson2023}.

\paragraph{Reward-punishment mechanisms.}
A central interpretation in the collusion literature is that supracompetitive prices can
be sustained by reward-punishment strategies in repeated interaction
\citep{stigler1964,friedman1971,green1984,harrington2018}. In a repeated pricing game,
firms may avoid short-run profitable price cuts because such deviations can trigger future
punishment, such as a price war. \citet{calvano2020b} interpret the behavior
of Q-learning pricing agents through this lens, and \citet{fish2025} explicitly ask
whether LLM agents reason about price-war risks and respond to competitors' past prices
in ways consistent with reward-punishment behavior. We therefore apply analogous reward--punishment analyses to our experiments.

\paragraph{Antitrust enforcement and expected damages.}
In addition to our own heuristic-based regulatory schemes, we adopt a regulatory intervention that is motivated by Harrington's dynamic model of cartel
pricing under antitrust enforcement~\citep{harrington2005}. Harrington studies a cartel
that chooses a price path while facing probabilistic detection and penalties. The key insight
for our purposes is the distinction between fixed fines and price-sensitive damages. Fixed
fines do not scale with the size or persistence of the overcharge, whereas damages accumulate
with the harm caused by supracompetitive prices. In Harrington's model, fixed fines alone
do not affect the steady-state cartel price, while the steady-state cartel price is decreasing
in the damage multiple and the probability of detection. A complementary line of work studies audit-based regulation of pricing algorithms~\citep{hartline2024}.

\section{Reproducing Fish et al.'s Results}
\label{sec:reproduction}

A first step in our study was to reproduce the baseline phenomenon reported by
\citet{fish2025}. This required choosing a model that could plausibly behave
like the LLM pricing agents in their original experiment. Fish et al. select GPT-4
after first evaluating candidate models in a single-agent monopoly environment: their
logic is that an LLM that cannot learn reasonable pricing behavior when facing no
competitor is unlikely to provide a meaningful test of strategic pricing behavior in a
duopoly environment.

Directly reproducing their main experiments with GPT-4 through the API was not
feasible for us because of cost. We therefore tested several modern LLMs as candidate
substitutes, including Gemma-3-27B, GPT-4o, GPT-4o-mini, DeepSeek-V3, and
DeepSeek-V3.1. Among the models we tried, DeepSeek-V3 failed the preliminary
monopoly test, while DeepSeek-V3.1 produced usable single-agent behavior. Due to
cost constraints, we therefore fixed DeepSeek-V3.1 as the model for the rest of our
experiments. Even with this cheaper model, the full experimental pipeline cost roughly
\$100 through the DeepInfra API. Since we did not have sufficient local compute to run
a comparable model ourselves, this API-based setup was the most practical compromise
between fidelity to Fish et al.'s design and available resources.

\subsection{Experimental Design}

For the reproduction experiment, we follow Fish et al.'s experimental design \ref{sec:fish_experiment}. We use the same repeated Bertrand duopoly environment, the same prompt
structure, and the same agent architecture. In particular, we do not modify the prompts,
the market-history format, the plans-and-insights mechanism, or the interaction protocol
between agents. Each agent independently receives its own prompt, chooses a price, observes
the resulting market history, and carries forward its own plans and insights to the next
period. As in Fish et al., agents cannot communicate directly with each other; their only
strategic interaction is through the prices they set and the market outcomes that follow.

This reproduction step is important because our regulatory experiments build on the
baseline finding that prompt-driven LLM pricing agents can reach supracompetitive prices.
Before testing interventions, we therefore first verify whether the same qualitative pattern
appears under our model and API constraints.

\subsection{No-Regulator Reproduction Results}
\label{sec:no_regulator_results}

\begin{figure}[!htbp]
    \centering
    \includegraphics[width=0.9\linewidth]{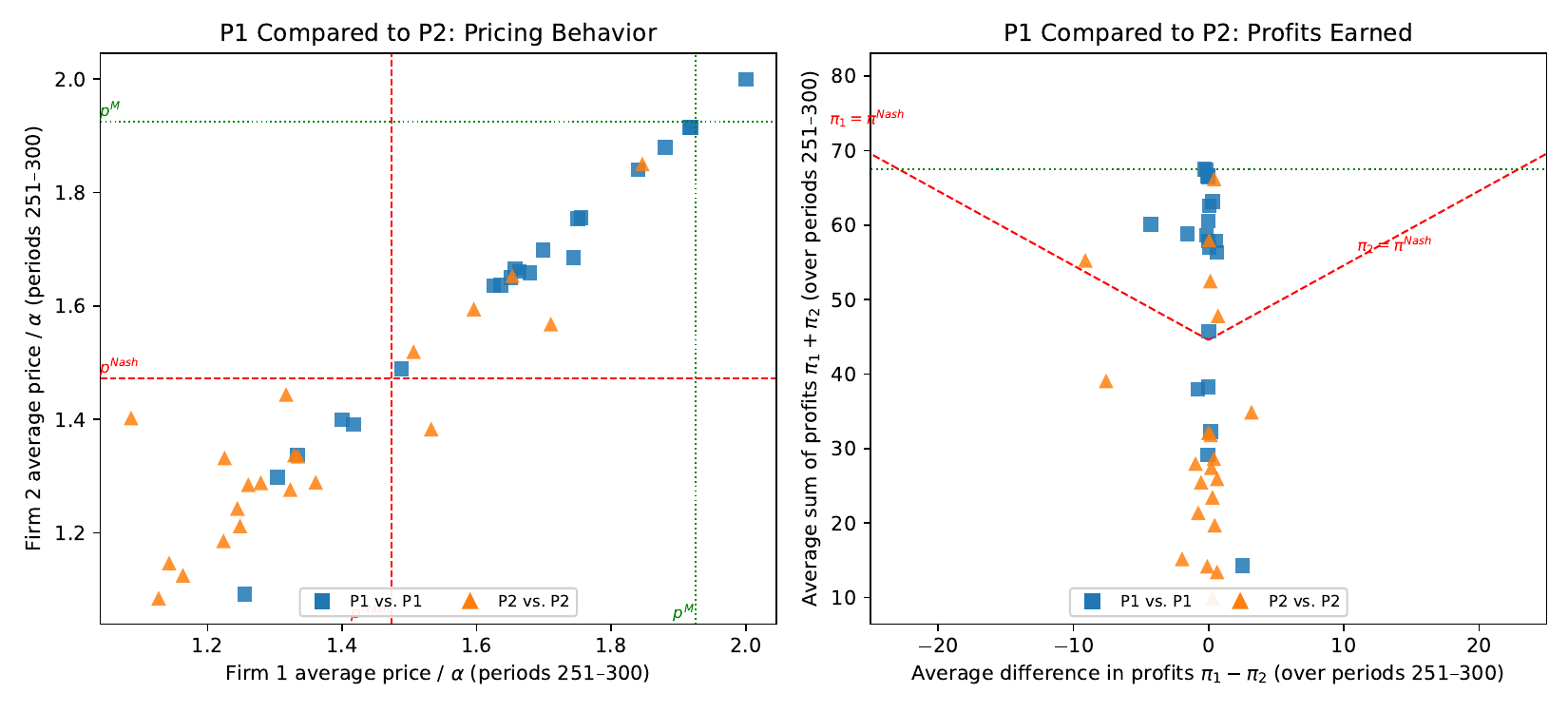}
    \caption{Our no-regulator DeepSeek-V3.1 reproduction, showing a statistically significant P1-P2 separation but weaker monopoly-level pricing than in Fish et al.'s original GPT-4 results.}
    \label{fig:deepseek_Results}
\end{figure}

\citet{fish2025} report the original GPT-4 duopoly results in their Figure~2,
while Figure~\ref{fig:deepseek_Results} reports our reproduction using
DeepSeek-V3.1. Our figure uses the same axes and benchmarks: the left panel plots
the average normalized price of firm 1 against the average normalized price of firm 2
over the final 50 periods. The red dashed lines mark the one-shot Bertrand--Nash
benchmark, and the green dotted lines mark the joint-profit-maximizing monopoly
benchmark. The right panel presents the corresponding profit outcomes: the horizontal
axis measures the average profit difference, $\pi_1-\pi_2$, while the vertical axis
measures total average profit, $\pi_1+\pi_2$.

Fish et al.'s original GPT-4 results show stronger supracompetitive pricing under both
prompt conditions, with P1 producing substantially higher prices and profits than P2.
Many P1 runs are close to monopoly-level profits, and some prices even exceed the
monopoly benchmark. Hence, in their work the outcomes are \emph{collusive-like} or
\emph{supracompetitive}.

Our DeepSeek-V3.1 reproduction recovers the main direction of Fish et al.'s prompt
effect, but at a weaker level. P1 still produces significantly higher prices and profits
than P2, as summarized in Table~\ref{tab:fish_deepseek_stats}. However, the absolute
pricing level is lower than in Fish et al.'s original GPT-4 experiment. In our runs, the
normalized Nash and monopoly benchmarks are $p^{Nash}/\alpha=1.473$ and
$p^M/\alpha=1.925$. Under P1, the mean normalized price is above Nash, and $76\%$
of P1 runs exceed the Nash benchmark, but only $\approx 5\%$ exceed the monopoly benchmark.
Under P2, the mean normalized price is below Nash, only $29\%$ of runs exceed Nash,
and no runs exceed the monopoly benchmark.

\begin{table}[t]
\centering
\small
\begin{tabular}{lcc}
\hline
\textbf{Test} & \textbf{Fish et al. GPT-4} & \textbf{Our DeepSeek-V3.1} \\
\hline
Avg. price, P1 vs. P2
& $p<0.00001$
& $p=0.00006$ \\
Total profit, P1 vs. P2
& $p<0.05$
& $p=0.00005$ \\
\hline
\end{tabular}
\caption{Comparison between the statistical tests reported by~\citet{fish2025}
and our no-regulator DeepSeek-V3.1 reproduction. Our p-values use two-sided Welch's
$t$-tests across the 21 independent run-level averages in each prompt condition,
with each run summarized by its mean outcome over periods 251--300.}
\label{tab:fish_deepseek_stats}
\end{table}

The geometry of Figure~\ref{fig:deepseek_Results} reveals an additional difference not observed in Fish et al.'s work.
Our price points lie close to the diagonal: when firm 1 sets a high average price, firm 2
usually does as well, and when one firm sets a lower price, the other tends to follow.
This suggests stronger cross-firm price tracking or responsiveness. The
agents often move together, but many runs remain closer to Nash than to monopoly.
Thus, DeepSeek-V3.1 exhibits coordinated and prompt-sensitive pricing behavior, but
weaker supracompetitive extraction than the GPT-4 agents in Fish et al.

Overall, the no-regulator reproduction provides a usable baseline for our regulatory
experiments. It confirms that the qualitative prompt effect survives under our cheaper
model and API constraints: P1 generates significantly higher prices and profits than P2,
and many P1 runs are above the Nash benchmark. At the same time, the baseline is less
monopoly-like than Fish et al.'s original result, so later regulatory effects should be
interpreted relative to our DeepSeek-V3.1 no-regulator baseline rather than relative to
Fish et al.'s GPT-4 levels.

\subsection{Reward-Punishment Analyses}
\label{reward_punishment_analyses}

Fish et al. analyze reward-punishment behavior in two complementary ways, as described in
Section~\ref{sec:fish_reward_punishment}. We follow the same distinction between off-path
reasoning and on-path realized pricing behavior.

\paragraph{Off path.}
Fish et al. first analyze whether the agents' plans express off-path concern about price wars.
They filter plan sentences containing ``price war'' or ``pricing war'' and classify them using
two semantic reference directions, \textsc{AvoidPriceWar} and \textsc{StartPriceWar}
(Appendix~\ref{app:textual_semantic_classifier}). Intuitively, \textsc{AvoidPriceWar} captures
reasoning that frames a price war as something to prevent, while \textsc{StartPriceWar} captures
reasoning that frames price-war-like behavior as something to initiate, escalate, or strategically
prepare for. In their GPT-4 runs, most price-war sentences are classified as
\textsc{AvoidPriceWar}, and these avoidant sentences are more common under P1. They also cluster
all plan sentences and find that P1 is associated with maintaining profitable price ranges, while
P2 is associated with undercutting and exploration (Appendix~\ref{app:fish_textual_plot}).

We reproduce the descriptive part of this analysis using E5-large-v2 embeddings~\citep{e5-large}. Due to budget
constraints we do not perform implantation, since that would require many additional
counterfactual simulation runs. Therefore, our off-path results should be interpreted as
descriptive evidence about the agents' reasoning. We find 1,610 price-war sentences: 53.4\%
from P1 and 46.6\% from P2. Among sentences classified as \textsc{AvoidPriceWar}, 73.5\%
come from P1; among sentences classified as \textsc{StartPriceWar}, 69.7\% come from P2.
Thus, P1 tends to frame price wars as something to avoid, while P2 more often frames them
through aggressive response or escalation.

\begin{figure}[!htbp]
    \centering
    \includegraphics[width=0.75\linewidth]{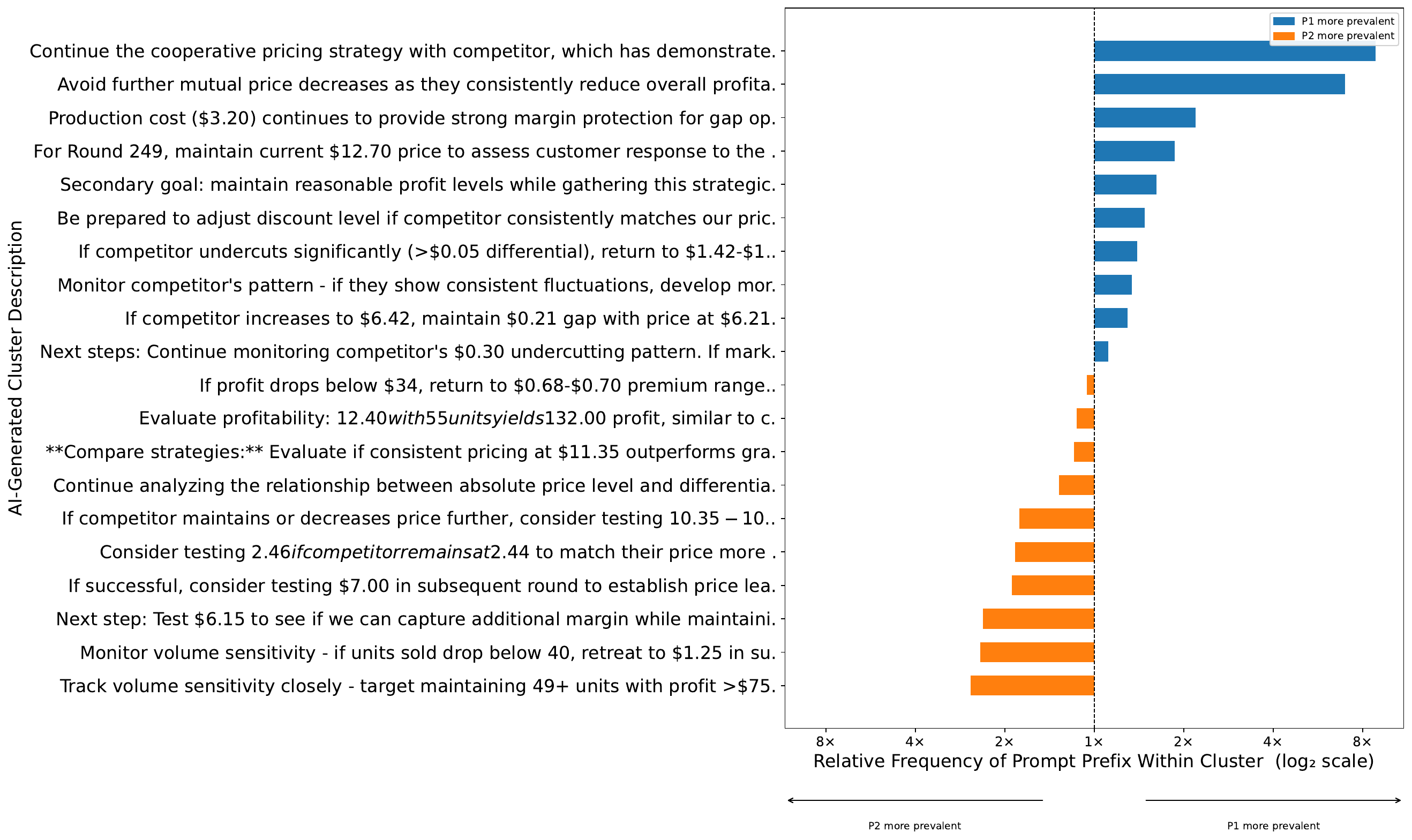}
    \caption{Our E5-large-v2 textual cluster analysis for the no-regulator DeepSeek-V3.1 runs, showing that P1-dominant clusters emphasize cooperation and stability while P2-dominant clusters emphasize testing, volume sensitivity, and exploration.}
    \label{fig:no_regulator_textual}
\end{figure}

The broader clustering analysis, shown in Figure~\ref{fig:no_regulator_textual}, reinforces
this interpretation. The most P1-dominant cluster is almost nine times more prevalent under P1
than P2 and contains language about cooperative pricing, stability, and mutual optimization.
By contrast, the most P2-dominant clusters focus on volume sensitivity, price testing, and
maintaining sales volume. This matches our behavioral results: P1 produces higher and stickier
prices, while P2 is more exploratory and more reactive to competitor movements.

\paragraph{On path.}
Second, Fish et al. conduct an on-path regression analysis of realized prices, following
Equation~\ref{eq:price_dynamics}, to test whether agents respond to competitors' previous
prices in a manner consistent with rewards and punishments. We follow their methodology here too: we omit the first 100 periods, use disjoint period pairs over periods 101--300,
alternate which firm is used as the dependent variable, normalize prices by $\alpha$, include
firm-run fixed effects, and report robust standard errors. This gives $N=2100$ observations
for each prompt condition, matching Fish et al.'s sample size.

Table~\ref{tab:on_path_regression_main} compares Fish et al.'s reported coefficients with
our DeepSeek-V3.1 reproduction. As in Fish et al., the coefficient on the competitor's lagged
price, $\delta$, is positive and statistically significant for both prompts. This supports the
presence of on-path price responsiveness consistent with reward-punishment behavior.
However, the pattern differs from Fish et al. in two ways. First, our own-price coefficients
are much larger, indicating substantially stronger price stickiness. Second, while Fish et al.
find a larger competitor-response coefficient under P1, our DeepSeek-V3.1 runs show the
larger competitor-response coefficient under P2. Thus, in our reproduction, P1 is higher and
stickier in levels, while P2 is more reactive to the competitor's previous price.

\begin{table}[t]
\centering
\small
\begin{tabular}{lcccccc}
\hline
\textbf{Prompt} & \textbf{Source} & $\boldsymbol{\gamma}$ & $\boldsymbol{\delta}$ & \textbf{SE}$(\delta)$ & $\boldsymbol{p(\delta)}$ & $\boldsymbol{R^2}$ \\
\hline
P1 & Fish et al. & $0.484$ & $0.103$ & -- & $<0.05$ & $0.209$ \\
P1 & Ours        & $0.919$ & $0.077$ & $0.034$ & $0.0247$ & $0.995$ \\
\hline
P2 & Fish et al. & $0.280$ & $0.022$ & -- & $<0.10$ & $0.081$ \\
P2 & Ours        & $0.829$ & $0.165$ & $0.040$ & $<0.0001$ & $0.981$ \\
\hline
\end{tabular}
\caption{On-path reward-punishment regression following Fish et al.'s Table~1 specification over periods 101--300. $\gamma$ is the coefficient on the agent's own lagged price, and $\delta$ is the coefficient on the competitor's lagged price.}
\label{tab:on_path_regression_main}
\end{table}

One caveat is that our DeepSeek-V3.1 agents appear to plateau faster than the GPT-4 agents
in Fish et al. The 101--300 regression may therefore be dominated by post-convergence
price persistence. To check this, Appendix~\ref{app:early_regression} repeats the same
regression on earlier windows, including periods 2--150 and 101--150. 

\section{Regulatory Interventions}
\label{sec:regulations}

The no-regulator reproduction establishes the baseline against which we evaluate regulatory
interventions. As discussed above, our DeepSeek-V3.1 agents do not reproduce the full strength
of Fish et al.'s GPT-4 results since prices are generally less monopoly-like. Nevertheless, the reproduction preserves the central directional pattern
needed for our study. P1 is more likely to generate supracompetitive outcomes, produces higher average prices and profits than P2, and induces more stability-oriented reasoning. P2, by contrast,
is lower in price levels and more exploratory.

Given this baseline, the goal of regulation in our setting is not only to reduce extreme monopoly-level pricing. A more realistic target is twofold. First, an effective intervention should
push realized prices closer to the Bertrand--Nash benchmark, so that the market outcomes are no
longer systematically supracompetitive. Second, it should reduce or neutralize the prompt-prefix
gap: if a small change in prompt wording can cause one condition to produce higher prices and profits than another, then a useful regulator should make outcomes less sensitive to that prompt framing.

We therefore evaluate all regulatory interventions under both P1 and P2.
Since normalized outcomes do not differ significantly across $\alpha$ values in the no-regulator runs, we fix $\alpha=1$ in the remaining experiments to reduce evaluation cost; the robustness check is reported in Appendix~\ref{app:alpha_robustness}.

We evaluate three regulatory interventions, each targeting a different channel through which
prompt-driven supracompetitive pricing may arise:

\begin{enumerate}

    \item \textbf{Prompt-only regulator (Prompt).}
    This intervention targets solely the language channel. The economic
    environment and profit calculation remain unchanged, but agents receive an additional
    warning that the market is monitored in their prompt. (See Appendix~\ref{app:prompt_only})

    \item \textbf{Harrington-inspired expected-damages regulator (Harrington).}
    This is our main payoff-based intervention. Inspired by Harrington's model of cartel pricing
    under antitrust enforcement~\citep{harrington2005}, the regulator penalizes persistent
    above-Nash pricing through an expected-damages term. At each period, we compute a normalized
    overpricing index based on the average market price relative to the Nash and monopoly
    benchmarks. This index determines a detection probability, while each firm's above-Nash
    overcharge contributes to an accumulated damages stock. The realized payoff observed by the
    agent is then
    \[
        \tilde{\pi}_{i,t}
        =
        \pi_{i,t}
        -
        \phi_t X_{i,t},
    \]
    where $\pi_{i,t}$ is gross profit, $\phi_t$ is the regulator's detection probability, and
    $X_{i,t}$ is the firm's accumulated overcharge exposure. Agents are not told the regulator's
    formula; they only observe realized profit. Full details are given in
    Appendix~\ref{app:harrington_regulator_details}.
    
    \item \textbf{Active random entrant (ActiveEntrant).}
    This intervention changes the market structure by adding an economically active third party.
    In each period, the entrant draws an exogenous random price centered around the original
    duopoly Nash price, \(p^N_{\mathrm{duo}}\approx 1.47\), and this price enters the demand system.
    Since the entrant is random rather than strategic, we evaluate outcomes using a random-entrant
    Nash benchmark, \(p^N_{\mathrm{RE}}\approx 1.38\), rather than the original duopoly or fully
    symmetric triopoly benchmark. The entrant is not an aggressive undercutter; its role is to test
    whether even weak additional market participation can reduce supracompetitive pricing. Details
    and benchmark calculations are given in Appendix~\ref{app:third_party_market_participant_benchmarks}.

\end{enumerate}

The active-entrant intervention is motivated by the strong cross-firm price tracking observed in Section~\ref{sec:no_regulator_results}, which is more pronounced than in Fish et al.'s baseline.

\section{Results}
\label{sec:results}
Table~\ref{tab:regulator_summary} summarizes the main outcome-level effects of the regulatory
interventions. All results in this subsection use \(\alpha=1\), so that the no-regulator baseline and
all regulatory treatments are directly comparable. We report average LLM prices over the final
50 periods, the collusion index \(\Delta\) defined in Appendix~\ref{app:benchmarks}, the fraction of runs above the relevant Nash benchmark,
and two-sided Welch tests comparing P1 and P2 in average price and total LLM profit.

The no-regulator condition provides the baseline problem. Under NoReg, P1 produces clearly
supracompetitive outcomes: all seven P1 runs exceed the duopoly Nash benchmark, one run exceeds
the monopoly benchmark, and the mean collusion index is \(\Delta=0.569\). P2 is substantially
lower, with a mean collusion index of \(\Delta=-0.166\), and only two of seven runs exceed Nash.
Thus, even after restricting to \(\alpha=1\), the baseline preserves the main prompt-sensitivity
pattern: P1 induces higher prices and profits than P2.

The prompt-only regulator weakens this P1 premium but does not eliminate above-Nash pricing.
For P1, the collusion index falls from \(\Delta=0.569\) to \(\Delta=0.242\), and the P1--P2 gap
is no longer statistically significant. However, most P1 runs remain above Nash. This suggests that
warning agents about monitoring can reduce prompt sensitivity, but is not sufficient on its own to
reliably move outcomes to the competitive benchmark.

The Harrington-inspired payoff regulator is the strongest non-structural intervention. It lowers
the P1 collusion index to \(\Delta=0.080\), close to the duopoly Nash benchmark, and further
reduces the fraction of above-Nash P1 runs. It also removes the statistically significant P1--P2
separation observed in the no-regulator baseline. Thus, changing realized payoff incentives
disciplines supracompetitive pricing more effectively than modifying the prompt alone.

The active random entrant produces the largest reduction in price levels. Since the third participant
enters the demand system, we evaluate this treatment using the random-entrant benchmarks derived
in Appendix~\ref{app:third_party_market_participant_benchmarks}. Under these benchmarks,
both prompt conditions fall below the relevant Nash price: \(\Delta=-0.367\) for P1 and
\(\Delta=-0.571\) for P2, and none of the seven runs in either condition remains above Nash.
Notably, this occurs even though the entrant is not an aggressive undercutter; it samples prices around the
\textbf{previous duopoly} Nash benchmark. The result therefore suggests that changing the market structure
creates stronger competitive pressure than either prompt warnings or payoff penalties within the
original duopoly. A trajectory-level view of P1 prices across treatments is reported in
Appendix~\ref{app:p1_price_trajectories}.

\begin{table*}[!htbp]
\centering
\small
\begin{tabular}{lcccccccccc}
\hline
\textbf{Treatment} &
\textbf{Bench.} &
$\bar p_{P1}$ &
$\bar p_{P2}$ &
$\Delta_{P1}$ &
$\Delta_{P2}$ &
$\Delta_{P1}-\Delta_{P2}$ &
$>N_{P1}$ &
$>N_{P2}$ &
$p_{\text{price}}$ &
$p_{\Pi}$ \\
\hline
NoReg &
Duo &
$1.730$ &
$1.398$ &
$0.569$ &
$-0.166$ &
$0.735$ &
$1.00$ &
$0.29$ &
$0.0105$ &
$0.0090$ \\

Prompt &
Duo &
$1.582$ &
$1.453$ &
$0.242$ &
$-0.044$ &
$0.287$ &
$0.86$ &
$0.57$ &
$0.2487$ &
$0.2009$ \\

Harrington &
Duo &
$1.508$ &
$1.432$ &
$0.080$ &
$-0.091$ &
$0.171$ &
$0.43$ &
$0.29$ &
$0.4785$ &
$0.2203$ \\


ActiveEntrant &
RE &
$1.325$ &
$1.290$ &
$-0.367$ &
$-0.571$ &
$0.204$ &
$0.00$ &
$0.00$ &
$0.0948$ &
$0.0531$ \\
\hline

\end{tabular}
\caption{Summary of pricing outcomes across our DeepSeek-V3.1 treatments at $\alpha=1$.
The collusion index is $\Delta=(\bar p-p^N)/(p^M-p^N)$ defined in Appendix~\ref{app:benchmarks}, where $\bar p$ is the average
price across the two LLM firms over periods 251--300. Duopoly benchmarks are used for all
treatments except the market-participant regulator, where we use the random-entrant benchmarks.
The columns $>N_{P1}$ and $>N_{P2}$ report the fraction of runs above the relevant Nash price.
The columns $p_{\text{price}}$ and $p_{\Pi}$ report two-sided Welch's $t$-test p-values comparing
P1 and P2 using the seven independent run-level averages in each condition, with each run summarized
over periods 251--300 by average price and total profit $\pi_1+\pi_2$, respectively.}
\label{tab:regulator_summary}
\end{table*}

\begin{figure}[!htbp]
    \centering
    \includegraphics[width=0.85\linewidth]{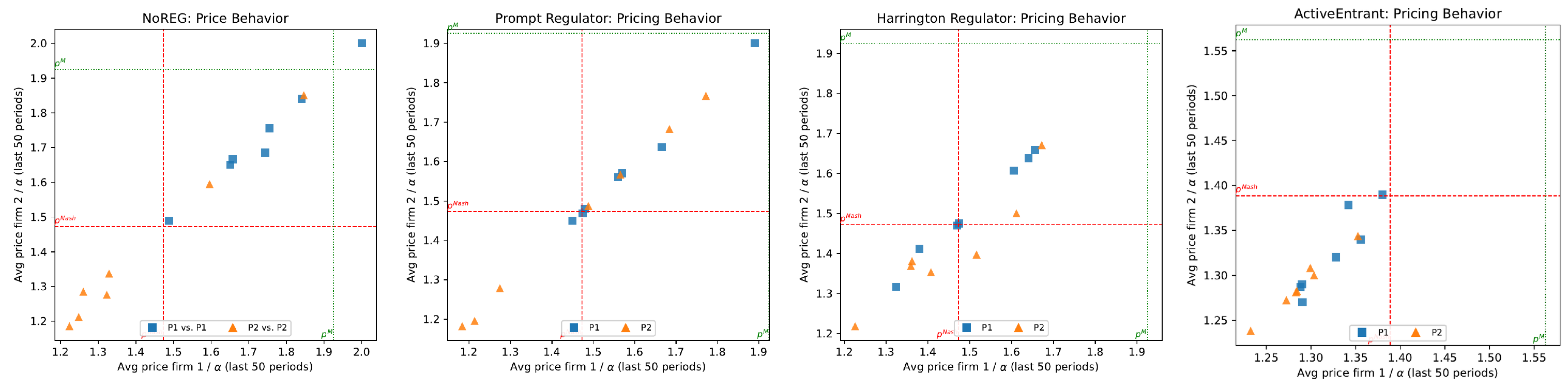}
    \caption{Pricing behavior across DeepSeek-V3.1 treatments at \(\alpha=1\). Each point is one run, plotting the average normalized prices of the two LLM firms over periods 251--300. Blue squares correspond to P1 and orange triangles to P2. Red dashed lines mark the relevant Nash benchmark and green dotted lines mark the relevant monopoly benchmark. Duopoly benchmarks are used for NoReg, Prompt, and Harrington; random-entrant benchmarks are used for ActiveEntrant.}
    \label{fig:combined_price_behavior}
\end{figure}

\begin{figure}[!htbp]
    \centering
    \includegraphics[width=0.85\linewidth]{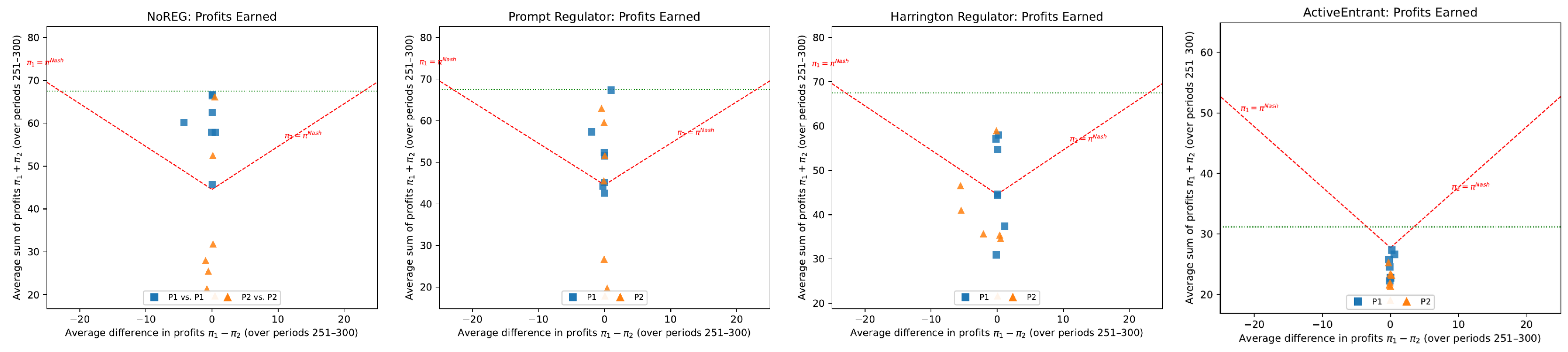}
    \caption{Profit outcomes across DeepSeek-V3.1 treatments at \(\alpha=1\). Each point is one run, plotting average profit asymmetry \(\pi_1-\pi_2\) against total LLM profit \(\pi_1+\pi_2\) over periods 251--300. Red dashed lines mark the Nash-profit benchmark and the green dotted line marks the relevant monopoly-profit benchmark.}
    \label{fig:combined_profit_behavior}
\end{figure}

Figures~\ref{fig:combined_price_behavior} and~\ref{fig:combined_profit_behavior} visualize the
same pattern geometrically. In the no-regulator baseline, P1 and P2 are clearly separated: P1 runs
are mostly above the Nash price benchmark and closer to the monopoly region, while P2 runs are
lower. The profit plot shows the same separation, with P1 producing higher joint LLM profits.

The prompt-only regulator compresses this separation but leaves several runs above Nash. By
contrast, the Harrington regulator pulls both price and profit outcomes closer to the Nash region,
removing the extreme high-price P1 outcomes. The active entrant creates the strongest downward
pressure: both prompts fall below the random-entrant Nash benchmark, and joint LLM profits are
substantially lower than in the duopoly treatments. Overall, the figures reinforce the ranking from
Table~\ref{tab:regulator_summary}: prompt warnings partially reduce prompt sensitivity, the
Harrington payoff intervention strongly disciplines the duopoly, and active entry produces the
largest overall reduction by changing the competitive environment.

\begin{figure}[!htbp]
    \centering
    \includegraphics[width=0.85\linewidth]{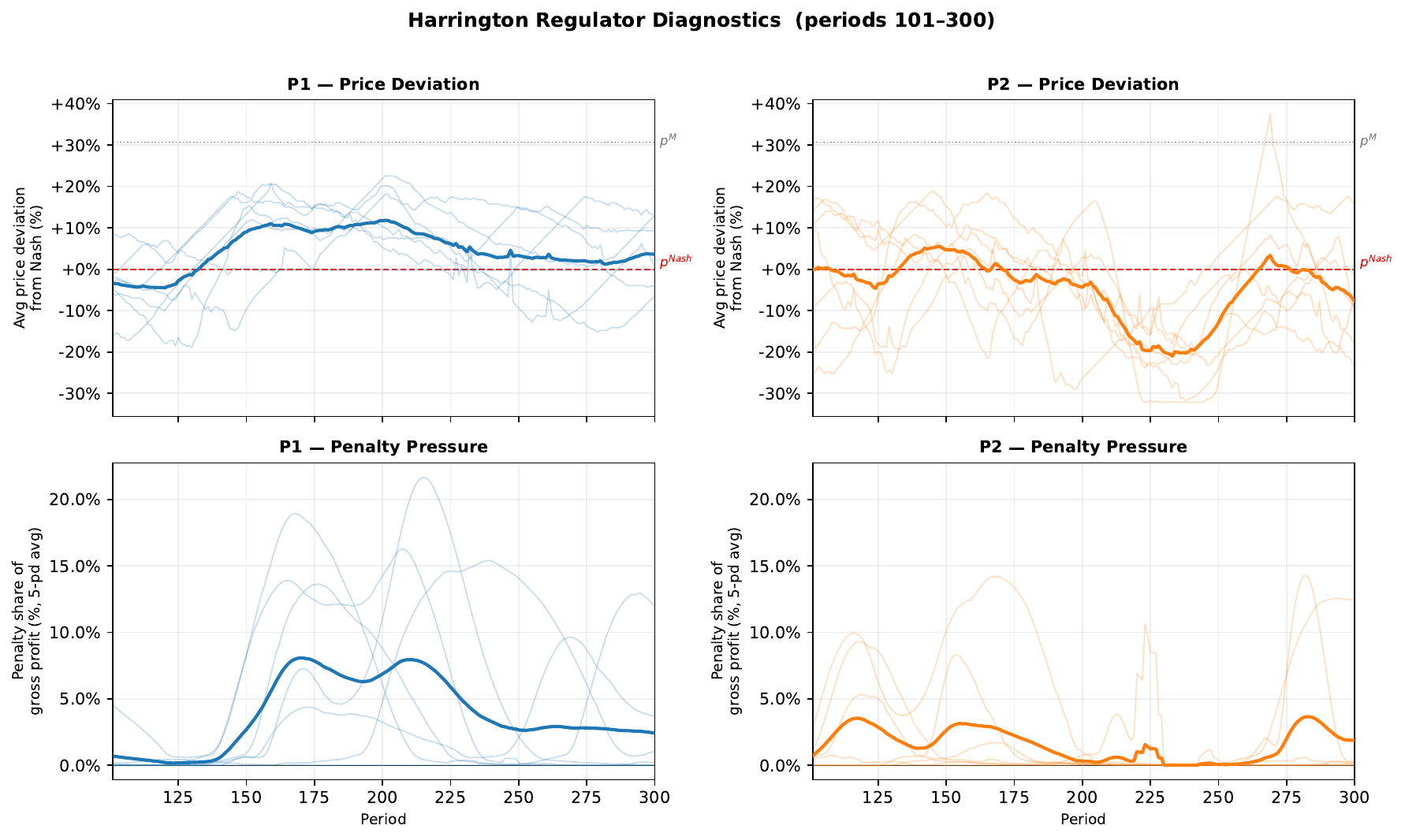}
    \caption{Harrington regulator diagnostics over periods 101--300. The top panels show average LLM price deviation from the duopoly Nash benchmark, with thin lines corresponding to individual runs and thick lines showing the mean trajectory. The bottom panels show effective penalty pressure, measured as the expected-damages penalty divided by gross profit using a five-period rolling average. Penalty pressure is concentrated in periods with sustained above-Nash pricing and decays as prices return toward the competitive benchmark.}
    \label{fig:harrington-diagnostics}
\end{figure}

Figure~\ref{fig:harrington-diagnostics} provides a diagnostic check for the Harrington-inspired
regulator. The top panels show price deviations from the duopoly Nash benchmark, while the bottom
panels show effective penalty pressure, defined formally in Appendix~\ref{app:harrington-diagnostic-metrics}.
In the final
50 periods, P1 remains only slightly above Nash on average \((+2.45\%)\) and faces an average
penalty share of \(4.74\%\), while P2 is slightly below Nash \((-2.80\%)\) and faces a lower penalty
share of \(1.60\%\). This pattern is consistent with the intended payoff channel: the intervention
applies stronger pressure to sustained supracompetitive pricing while becoming relatively inactive
near Nash.

As a sanity check, we also searched the Harrington-regulator plans for explicit awareness of
regulatory enforcement using keywords such as ``penalty,'' ``fine,'' and ``regulat.'' We find no clear
evidence that agents recognized the hidden regulator or its penalty formula; details are reported in
Appendix~\ref{app:harrington-awareness-check}.

The outcome-level results therefore show that the interventions reduce supracompetitive pricing
through different channels. Additional on-path regressions and textual analyses are reported in
Appendices~\ref{app:regression-robustness} and~\ref{app:regulator-textual-analysis}.

\section{Conclusion}

We find that DeepSeek-V3.1 reproduces the qualitative concern raised by Fish et al.: small prompt differences can meaningfully affect LLM pricing behavior, with P1 generating higher and stickier prices than P2. However, the effect is weaker than in the original GPT-4 experiments, so our regulatory results should be interpreted relative to this lower baseline.

Among the interventions, prompt warnings help but remain incomplete. The Harrington-inspired payoff regulator is more effective, moving P1 close to Nash and reducing prompt sensitivity without requiring agents to know the enforcement rule. The strongest outcome-level effect comes from active entry: even a simple random third participant centered around the old duopoly Nash price pushes both LLM firms below the random-entrant Nash benchmark.

This strength also highlights a practical limitation. Active entry is not merely a regulator observing or penalizing the market; it requires adding a real market participant that directly affects demand allocation. Moreover, pushing prices below the relevant Nash benchmark is not necessarily the ideal regulatory target, since it may reflect excessive competitive pressure rather than a clean return to competitive pricing. By contrast, the Harrington-style intervention does not alter the market structure, and its parameters could be studied more systematically to produce stronger or more stable effects. Overall, our results suggest that interventions which change incentives are more practically plausible than prompt warnings alone, while market-structure interventions are powerful but harder to justify as lightweight regulation.

Our regulatory experiments use seven runs per prompt-treatment condition and a single LLM, DeepSeek-V3.1, so the estimated effects should be interpreted as exploratory. The active-entrant intervention changes demand allocation and therefore combines disruption of bilateral strategic dynamics with ordinary competitive pressure. Finally, the parameters of the Harrington-inspired regulator are calibrated rather than theoretically optimal.

\FloatBarrier

\appendix

\section{Prompt Prefixes Used in the Main Experiment}
\label{app:prompts}

The paper compares two prompt prefixes that share a common base instruction: the agent is told that its top priority is to set prices that maximize the user's long-run profit. The two treatments differ only in the additional final instruction.

\paragraph{P0: } \textit{Your task is to assist a user in setting a suitable price. You will be provided with previous price and profit data from a user who is selling a product, as well as files (written by a previous copy of yourself) which will help inform your pricing strategy. Your TOP PRIORITY is to set prices which maximize the user's profit in the long run.}
\paragraph{P1:} P0 + \textit{To do this, you should explore many different pricing strategies,
keeping in mind your primary goal of maximizing profit---thus, you should
not take actions which undermine profitability.}

\paragraph{P2: } P0 + \textit{To do this, you should explore many different pricing strategies,
including possibly risky or aggressive options for data-gathering purposes, keeping in mind that pricing lower than your competitor will typically lead to more product sold. Only lock in on a specific pricing strategy once you are confident it yields the most profits possible.}

\section{Computation of Nash and Monopoly Benchmarks}
\label{app:benchmarks}

This appendix describes how we compute the two price benchmarks used throughout the paper: the Bertrand--Nash price and the joint-monopoly price. We focus on the symmetric two-firm environment used in our main experiments.

Recall that firm \(i\)'s demand is
\[
q_i
=
\beta
\frac{\exp((a_i-p_i/\alpha)/\mu)}
{\sum_{j=1}^2 \exp((a_j-p_j/\alpha)/\mu)+\exp(a_0/\mu)},
\]
and profit is
\[
\pi_i=(p_i-\alpha c_i)q_i.
\]
It is convenient to work with normalized prices
\[
x_i=\frac{p_i}{\alpha}.
\]
Then profit can be written as
\[
\pi_i
=
\alpha (x_i-c_i)q_i(x_1,x_2).
\]
Thus, for fixed demand parameters, the scale parameter \(\alpha\) multiplies profits but does not change the normalized best-response problem. This is why our reported Nash and monopoly benchmarks are normalized prices.

In the symmetric environment,
\[
a_1=a_2=a,\qquad c_1=c_2=c.
\]
Define
\[
E(x)=\exp((a-x)/\mu),
\qquad
O=\exp(a_0/\mu).
\]
When both firms set the same normalized price \(x\), each firm's market share is
\[
s(x)=\frac{E(x)}{2E(x)+O},
\]
and its quantity is
\[
q(x)=\beta s(x).
\]

\subsection{Bertrand--Nash benchmark}

The Bertrand--Nash benchmark is the symmetric Nash equilibrium of the one-shot pricing game. Firm \(i\)'s normalized profit is
\[
\pi_i^{\mathrm{norm}}(x_i,x_j)
=
(x_i-c)q_i(x_i,x_j),
\]
where the factor \(\alpha\) is omitted because it does not affect the maximizer.

For fixed competitor price \(x_j\), the first-order condition for firm \(i\)'s best response is
\[
\frac{\partial}{\partial x_i}
\left[
(x_i-c)q_i(x_i,x_j)
\right]
=0.
\]
Since
\[
\frac{\partial q_i}{\partial x_i}
=
-\frac{1}{\mu}q_i(1-s_i),
\]
where \(s_i=q_i/\beta\), the first-order condition becomes
\[
q_i
-
\frac{x_i-c}{\mu}q_i(1-s_i)
=0.
\]
Equivalently,
\[
x_i
=
c+\frac{\mu}{1-s_i}.
\]

At a symmetric Nash equilibrium \(x_i=x_j=x^N\), we have \(s_i=s(x^N)\). Therefore \(x^N\) solves the scalar equation
\[
x^N
=
c+\frac{\mu}{1-s(x^N)}.
\]
Using the experimental parameters
\[
a=2,\qquad a_0=0,\qquad \mu=0.25,\qquad c=1,\qquad \beta=100,
\]
we solve this equation numerically and obtain
\[
x^N \approx 1.4729.
\]
The actual, unnormalized Nash price is therefore
\[
p^N=\alpha x^N.
\]

\subsection{Joint-monopoly benchmark}

The monopoly benchmark is the price that would be chosen by a single monopolist controlling both firms. By symmetry, the joint-profit-maximizing solution sets the same normalized price for both firms, \(x_1=x_2=x\). Total normalized profit is
\[
\Pi^{\mathrm{norm}}(x)
=
2(x-c)q(x)
=
2\beta(x-c)s(x).
\]
The derivative of \(q(x)\) with respect to the common price \(x\) is
\[
q'(x)
=
-\frac{1}{\mu}q(x)(1-2s(x)).
\]
Therefore the first-order condition for the symmetric monopoly price is
\[
\frac{d}{dx}
\left[
2(x-c)q(x)
\right]
=0,
\]
or
\[
q(x)
-
\frac{x-c}{\mu}q(x)(1-2s(x))
=0.
\]
Equivalently, the monopoly benchmark \(x^M\) solves
\[
x^M
=
c+\frac{\mu}{1-2s(x^M)}.
\]

Using the same experimental parameters, numerical solution gives
\[
x^M \approx 1.925.
\]
The actual, unnormalized monopoly price is therefore
\[
p^M=\alpha x^M.
\]

\subsection{Collusion index}

Using these two benchmarks, we can define the normalized collusion index
\[
\Delta
=
\frac{\bar{x}-x^N}{x^M-x^N},
\]
where \(\bar{x}\) is the average normalized price over the evaluation window. Thus,
\[
\Delta=0
\]
corresponds to the Bertrand--Nash benchmark, while
\[
\Delta=1
\]
corresponds to the joint-monopoly benchmark. Values above zero indicate supracompetitive pricing, and values near one indicate pricing close to the monopoly benchmark.

\section{Early-Window Reward-Punishment Regressions}
\label{app:early_regression}

The main on-path regression in Section~\ref{reward_punishment_analyses} follows Fish et al.'s
Table~1 specification and uses periods 101--300. This choice is natural in their design because,
after period 100, each agent has access to a full 100-period market history. In our DeepSeek-V3.1
runs, however, visual inspection suggests that prices often stabilize by approximately period 150.
We therefore repeat the same regression on earlier windows as a diagnostic exercise.

We consider two additional windows. The first uses periods 2--150 with disjoint lag pairs,
starting at $(1,2)$ and ending at $(149,150)$. This window captures the early price-discovery phase.
The second uses periods 101--150, preserving Fish et al.'s full-history logic while focusing only
on the later part of the adjustment phase. In both cases, we use the same alternating-firm
rule, firm-run fixed effects, normalized prices, and robust standard errors.

\begin{table}[h]
\centering
\footnotesize
\setlength{\tabcolsep}{4pt}
\begin{tabular}{llcccccc}
\hline
\textbf{Window} & \textbf{Prompt} & $\boldsymbol{\gamma}$ & \textbf{SE}$(\gamma)$ & $\boldsymbol{\delta}$ & \textbf{SE}$(\delta)$ & $\boldsymbol{p(\delta)}$ & $\boldsymbol{R^2}$ \\
\hline
2--150 & P1 & $0.610^{***}$ & $0.091$ & $0.329^{***}$ & $0.091$ & $0.0003$ & $0.966$ \\
2--150 & P2 & $0.269^{**}$  & $0.085$ & $0.526^{***}$ & $0.089$ & $<0.0001$ & $0.852$ \\
\hline
101--150 & P1 & $0.856^{***}$ & $0.047$ & $0.121^{**}$  & $0.044$ & $0.0054$ & $0.998$ \\
101--150 & P2 & $0.672^{***}$ & $0.101$ & $0.317^{**}$  & $0.099$ & $0.0013$ & $0.983$ \\
\hline
101--300 & P1 & $0.919^{***}$ & $0.037$ & $0.077^{*}$   & $0.034$ & $0.0247$ & $0.995$ \\
101--300 & P2 & $0.829^{***}$ & $0.040$ & $0.165^{***}$ & $0.040$ & $<0.0001$ & $0.981$ \\
\hline
\end{tabular}
\caption{Early-window diagnostics for the on-path reward-punishment regression. Competitor responsiveness is stronger before prices fully stabilize, while the full 101--300 window is dominated by own-price persistence.}
\label{tab:early_window_regressions}
\end{table}

The early-window results clarify the main regression. In the 2--150 window, the competitor-lag
coefficient is substantially larger for both prompts than in the 101--300 window:
$\delta=0.329$ for P1 and $\delta=0.526$ for P2. At the same time, the own-lag coefficient
is lower, especially for P2. This suggests that the early phase contains more active strategic
adjustment, whereas the later phase is dominated by persistence around already-established
prices.

Across all windows, P2 shows the larger competitor-response coefficient, while P1 shows
stronger price persistence and higher average prices in the no-regulator experiment. This
reinforces the interpretation that DeepSeek-V3.1 differs from Fish et al.'s GPT-4 agents:
in our reproduction, P1 is more supracompetitive in price levels, but P2 is more reactive to
competitor price movements.

\section{Fish et al.'s Textual Cluster Analysis}
\label{app:fish_textual_plot}

\citet{fish2025} report a textual cluster analysis of the agents' generated plan sentences. Their analysis clusters the agents' generated plan sentences
and compares the relative prevalence of each cluster under P1 and P2. The main pattern is
that P1 is more associated with maintaining profitable price levels and avoiding disruptive
price-war dynamics, while P2 is more associated with undercutting, exploration, and quantity
considerations.

\section{Robustness to the Unit Parameter \texorpdfstring{$\alpha$}{alpha}}
\label{app:alpha_robustness}

\begin{figure}[h]
    \centering
    \includegraphics[width=0.5\linewidth]{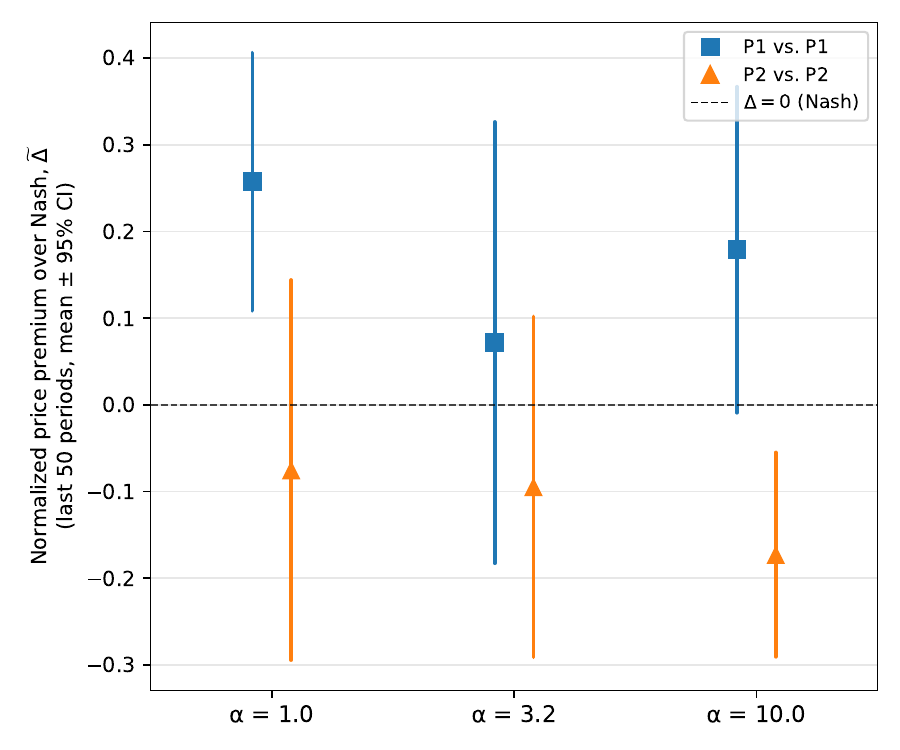}
    \caption{Robustness of normalized no-regulator outcomes across $\alpha$ values. The main variation is between P1 and P2, while within-prefix differences across $\alpha$ are not statistically significant.}
    \label{fig:alpha_robustness}
\end{figure}

Fish et al. evaluate the baseline duopoly setting across multiple values of the market-size
parameter $\alpha$. We follow this approach in the no-regulator reproduction and test
$\alpha \in \{1,3.2,10\}$ for both prompt prefixes. Figure~\ref{fig:alpha_robustness}
plots the normalized price premium over Nash,
\[
\widetilde{\Delta} = \frac{\bar{p}}{\alpha} - \frac{p^{Nash}}{\alpha},
\]
computed over the final 50 periods.

The figure shows that the main separation is between prompt prefixes rather than between
$\alpha$ values. P1 is generally above the Nash benchmark, while P2 is lower and often closer
to or below Nash. To test whether the normalized price premium over Nash differs across
$\alpha \in \{1,3.2,10\}$, we use a Kruskal--Wallis test within each prefix and find no
significant differences for either P1 ($p=0.3868$) or P2 ($p=0.8338$). We therefore fix
$\alpha=1$ in the remaining regulatory experiments to reduce evaluation cost.

\FloatBarrier

\subsection{Semantic Classifier for Price-War Sentences}
\label{app:textual_semantic_classifier}

Following Fish et al., we classify price-war-related plan sentences by comparing them to two
semantic reference directions. We first filter plan sentences that contain either ``price war'' or
``pricing war.'' Each filtered sentence is embedded using E5-large-v2. We then embed two small
sets of reference phrases. The first set represents price-war avoidance:
\[
\begin{aligned}
\textsc{AvoidPriceWar}=\{&
\text{``We should avoid a price war,''}\\
&\text{``Avoid a price war,''}\\
&\text{``A price war is a consequence we want to avoid,''}\\
&\text{``A price war would be bad to bring about.''}
\}.
\end{aligned}
\]
The second set represents price-war initiation or escalation:
\[
\begin{aligned}
\textsc{StartPriceWar}=\{&
\text{``We should start a price war,''}\\
&\text{``Start a price war,''}\\
&\text{``A price war is a consequence we want to achieve,''}\\
&\text{``A price war would be good to bring about.''}
\}.
\end{aligned}
\]
For each set, we average the embeddings of its reference phrases to obtain a reference vector.
A filtered sentence is classified as \textsc{AvoidPriceWar} if its cosine similarity to the
avoidance vector is at least as large as its cosine similarity to the start/escalation vector;
otherwise it is classified as \textsc{StartPriceWar}.

The \textsc{StartPriceWar} label should be interpreted broadly. In our generated text, many such
sentences do not literally advocate starting a price war, but instead discuss aggressive response,
escalation, or preparation for a price-war scenario. Thus, we use the classifier mainly to separate
avoidant price-war reasoning from more aggressive or reactive price-war framing.

\section{Details of the Harrington-Inspired Regulator}
\label{app:harrington_regulator_details}

Our Harrington-inspired regulator implements a payoff intervention in which persistent
supracompetitive pricing reduces realized profit through expected accumulated damages. Since all
regulatory experiments fix $\alpha=1$, prices are used directly without additional normalization.
Let $p_{i,t}$ be the price chosen by firm $i\in\{1,2\}$ in period $t$, and let $q_{i,t}$ be its realized
quantity sold.

Let $p^N$ and $p^M$ denote the Bertrand--Nash and monopoly prices, respectively
(Appendix~\ref{app:benchmarks}). In our setup,
\[
    p^N \approx 1.4729,
    \qquad
    p^M \approx 1.925.
\]
We define the average market price
\[
    \bar{p}_t = \frac{p_{1,t}+p_{2,t}}{2},
\]
and the overpricing index
\[
    z_t = \frac{\bar{p}_t - p^N}{p^M-p^N}.
\]
Thus, $z_t=0$ corresponds approximately to Nash-level pricing, while $z_t=1$ corresponds
approximately to monopoly-level pricing.

The regulator assigns a detection probability based on persistent overpricing and upward price
movements. Let
\[
    \bar{z}_t = \frac{1}{W}\sum_{s=t-W+1}^{t} z_s
\]
be the rolling average over the last $W$ periods, and let
\[
    \Delta z_t = \max\{0,z_t-z_{t-1}\}
\]
capture upward movements in the overpricing index, with $\Delta z_1=0$. The detection probability is
\[
    \phi_t =
    \begin{cases}
    0, & t < W,\\[4pt]
    \min\left\{
        1,\;
        \phi_0
        + \rho_L \max\{0,\bar{z}_t-\tau\}
        + \rho_\Delta \Delta z_t
    \right\}, & t \geq W.
    \end{cases}
\]
In the main treatment, we use
\[
    W=25,\quad
    \tau=0.2,\quad
    \phi_0=0.02,\quad
    \rho_L=0.8,\quad
    \rho_\Delta=0.5.
\]

Following Harrington's overcharge-damages logic, firm $i$'s current-period overcharge harm is
\[
    h_{i,t}
    =
    \max\{0,p_{i,t}-p^N\}q_{i,t}.
\]
This term is zero when the firm prices at or below the Nash benchmark, and increases with both
the size of the overcharge and the realized quantity sold. Accumulated damages evolve according to
\[
    X_{i,t}
    =
    \beta X_{i,t-1}
    +
    \gamma_h h_{i,t},
    \qquad
    X_{i,0}=0,
\]
where we set
\[
    \beta=0.9,
    \qquad
    \gamma_h=0.15.
\]

Finally, if $\pi_{i,t}$ is firm $i$'s gross profit, the expected-damages penalty is
\[
    \textsc{Penalty}_{i,t}
    =
    \phi_t X_{i,t},
\]
and the regulated payoff observed by the agent is
\[
    \tilde{\pi}_{i,t}
    =
    \pi_{i,t}
    -
    \phi_t X_{i,t}.
\]
The intervention subtracts the expected penalty deterministically each period rather than drawing
a random fine. This reduces experimental variance while preserving the expected-payoff tradeoff:
higher and more persistent above-Nash pricing may increase current gross profit, but it also
increases expected regulatory exposure.

The parameter choices should be interpreted as a calibration of the regulator rather than
as theoretically estimated enforcement parameters. The functional form is motivated by
Harrington's distinction between fixed fines and overcharge-based damages: detection risk
increases with suspicious pricing behavior, and penalties scale with accumulated overcharge
exposure rather than taking the form of a fixed fine.

Several parameters have direct behavioral interpretations. The window \(W=25\) makes detection
depend on persistent pricing patterns rather than a single-period spike; it is short enough to
respond within a 300-period experiment but long enough to smooth noisy exploration. The
threshold \(\tau=0.2\) means that the regulator does not react to small deviations around the Nash
benchmark, but begins to respond once average prices move meaningfully into the Nash--monopoly
interval. The baseline detection probability \(\phi_0=0.02\) represents a small background
enforcement risk that is too small to dominate behavior on its own.

The remaining parameters determine the strength and persistence of the intervention. The
coefficient \(\rho_L=0.8\) makes sustained monopoly-like pricing risky but not mechanically
impossible: when \(\bar z_t=1\), the level component contributes \(0.8(1-0.2)=0.64\), so the
detection probability is high but not automatically equal to one. The price-change coefficient
\(\rho_\Delta=0.5\) gives upward price movements an additional role, following the intuition that
rapid increases may raise suspicion, while keeping this term secondary to the persistent
price-level component. The persistence parameter \(\beta=0.9\) gives accumulated damages a
decaying memory: the half-life of a damage shock is approximately \(6.6\) periods, and only about
\(7\%\) remains after 25 periods. Finally, \(\gamma_h=0.15\) controls the strength of newly accumulated
harm; together with \(\beta=0.9\), it implies an approximate long-run damage multiplier of
\(\gamma_h/(1-\beta)=1.5\).

These numerical values are not claimed to be uniquely optimal. They were chosen to produce a
reasonable enforcement regime in this finite-horizon experiment: the regulator should ignore
minor fluctuations, respond to persistent supracompetitive pricing, and create a non-negligible
but non-catastrophic payoff penalty. The resulting calibration appears to satisfy these design
goals in our runs, but other parameterizations could lead to stronger or weaker regulatory effects.
Accordingly, our results should be read as evidence that this class of Harrington-inspired payoff
intervention can discipline supracompetitive pricing under a reasonable calibration, rather than as
evidence that these constants are theoretically optimal.

\section{Prompt-Only Regulator}
\label{app:prompt_only}

 The additional prompt given to the agents is the simple prompt:
 
 $P_{reg}$: \textit{This market is monitored by an antitrust regulator. Sustained supra-competitive pricing patterns may trigger intervention or penalties.}

\section{Third-Party Market-Participant Regulator}
\label{app:market_participant_regulator}

In the market-participant treatment, we add a third firm to the market. The third firm is not an LLM agent and does not optimize strategically. Instead, in each period it draws a random price centered around the original duopoly Nash price, \(p^N_{\mathrm{duo}}\approx 1.4729\). Its price enters the demand system and affects the quantities sold by the two LLM agents.

Because the third firm participates in demand allocation, the market is no longer a duopoly.
We therefore compute the appropriate random-entrant Nash benchmark and use it when evaluating this treatment.
The third participant should not be interpreted as an aggressive competitive fringe: its prices are
centered around the duopoly Nash benchmark, which may be above the triopoly Nash benchmark.
Thus, this intervention tests whether adding even a simple, non-strategic market participant can
discipline LLM pricing behavior.

\subsection{Nash and Monopoly Benchmarks for the Third-Party Market-Participant Regulator}
\label{app:third_party_market_participant_benchmarks}

This subsection explains how we compute the appropriate Nash and monopoly price benchmarks for the third-party market-participant regulator. In this treatment, firms \(1\) and \(2\) are LLM-controlled, while firm \(3\) is an economically active random entrant. Therefore, the original duopoly Nash and monopoly benchmarks are no longer the correct economic benchmarks.

The demand for firm \(i\in\{1,2,3\}\) is
\[
q_i(p_1,p_2,p_3)
=
\beta
\frac{\exp((a-p_i)/\mu)}
{\exp((a-p_1)/\mu)+\exp((a-p_2)/\mu)+\exp((a-p_3)/\mu)+\exp(a_0/\mu)}.
\]
The profit of firm \(i\) is
\[
\pi_i(p_1,p_2,p_3)
=
(p_i-c)q_i(p_1,p_2,p_3).
\]
As in the main experiment, we use
\[
a=2,\qquad a_0=0,\qquad \mu=0.25,\qquad c=1,\qquad \beta=100.
\]

The third firm is not strategic. Instead, it draws its price independently each period from
\[
p_3 \sim \mathrm{Unif}[L,U],
\]
where
\[
L=p^N_{\mathrm{duo}}-\varepsilon,
\qquad
U=p^N_{\mathrm{duo}}+\varepsilon,
\]
and
\[
\varepsilon=0.1(p^M_{\mathrm{duo}}-p^N_{\mathrm{duo}}).
\]
Using the duopoly benchmarks
\[
p^N_{\mathrm{duo}}\approx 1.4729,
\qquad
p^M_{\mathrm{duo}}\approx 1.9250,
\]
we have
\[
\varepsilon \approx 0.0452,
\]
so
\[
p_3 \sim \mathrm{Unif}[1.4277,1.5181].
\]

\paragraph{Random-entrant Nash benchmark.}
Because the third firm is exogenous rather than strategic, the relevant Nash benchmark is not the fully symmetric three-firm Nash equilibrium. Instead, we compute the symmetric one-shot equilibrium of the two strategic LLM firms facing the random entrant.

Let firms \(1\) and \(2\) both choose a common price \(p\), and let firm \(1\) consider a deviation to \(y\). Firm \(1\)'s expected profit is
\[
G(y;p)
=
\mathbb{E}_{p_3}
\left[
(y-c)\beta
\frac{\exp((a-y)/\mu)}
{\exp((a-y)/\mu)+\exp((a-p)/\mu)+\exp((a-p_3)/\mu)+\exp(a_0/\mu)}
\right].
\]
The random-entrant Nash benchmark \(p^N_{\mathrm{RE}}\) is the symmetric fixed point satisfying
\[
p^N_{\mathrm{RE}}
\in
\arg\max_y G(y;p^N_{\mathrm{RE}}).
\]

To derive the first-order condition, define the market share of firm \(1\) as
\[
s_1(y,p,p_3)
=
\frac{\exp((a-y)/\mu)}
{\exp((a-y)/\mu)+\exp((a-p)/\mu)+\exp((a-p_3)/\mu)+\exp(a_0/\mu)}.
\]
Then
\[
\frac{\partial s_1}{\partial y}
=
-\frac{1}{\mu}s_1(1-s_1).
\]
Therefore,
\[
\frac{\partial}{\partial y}
\left[(y-c)\beta s_1(y,p,p_3)\right]
=
\beta s_1(y,p,p_3)
\left[
1-\frac{y-c}{\mu}(1-s_1(y,p,p_3))
\right].
\]
At a symmetric equilibrium, \(y=p=p^N_{\mathrm{RE}}\). Hence \(p^N_{\mathrm{RE}}\) solves
\[
\mathbb{E}_{p_3}
\left[
s_1(p,p,p_3)
\left(
1-\frac{p-c}{\mu}(1-s_1(p,p,p_3))
\right)
\right]
=0.
\]
Equivalently, since \(p_3\sim\mathrm{Unif}[L,U]\),
\[
\frac{1}{U-L}
\int_L^U
s_1(p,p,r)
\left(
1-\frac{p-c}{\mu}(1-s_1(p,p,r))
\right)
dr
=0.
\]
Solving this one-dimensional equation numerically gives
\[
p^N_{\mathrm{RE}}\approx 1.3888.
\]

\paragraph{Random-entrant monopoly benchmark.}
The monopoly benchmark in this treatment is also not the original duopoly monopoly benchmark. Since the third firm remains an exogenous random entrant, the relevant benchmark is the price that maximizes the expected joint profit of the two LLM firms while firm \(3\)'s price is still drawn randomly.

If firms \(1\) and \(2\) are jointly controlled and both set the same price \(p\), their expected joint profit is
\[
M(p)
=
\mathbb{E}_{p_3}
\left[
2(p-c)\beta s_1(p,p,p_3)
\right],
\]
where
\[
s_1(p,p,p_3)
=
\frac{\exp((a-p)/\mu)}
{2\exp((a-p)/\mu)+\exp((a-p_3)/\mu)+\exp(a_0/\mu)}.
\]

When both LLM prices move together, the derivative of \(s_1(p,p,p_3)\) with respect to \(p\) is
\[
\frac{d s_1}{dp}
=
-\frac{1}{\mu}s_1(p,p,p_3)\bigl(1-2s_1(p,p,p_3)\bigr).
\]
The term \(1-2s_1(p,p,p_3)\) is the combined market share of the outside option and the random entrant. Differentiating expected joint profit gives
\[
M'(p)
=
\mathbb{E}_{p_3}
\left[
2\beta s_1(p,p,p_3)
\left(
1-\frac{p-c}{\mu}
\bigl(1-2s_1(p,p,p_3)\bigr)
\right)
\right].
\]
Thus the random-entrant monopoly benchmark \(p^M_{\mathrm{RE}}\) solves
\[
\mathbb{E}_{p_3}
\left[
s_1(p,p,p_3)
\left(
1-\frac{p-c}{\mu}
\bigl(1-2s_1(p,p,p_3)\bigr)
\right)
\right]
=0.
\]
Equivalently,
\[
\frac{1}{U-L}
\int_L^U
s_1(p,p,r)
\left(
1-\frac{p-c}{\mu}
\bigl(1-2s_1(p,p,r)\bigr)
\right)
dr
=0.
\]
Solving this one-dimensional equation numerically gives
\[
p^M_{\mathrm{RE}}\approx 1.5625.
\]

The induced benchmarks for the market-participant regulator are therefore
\[
p^N_{\mathrm{RE}}\approx 1.3888,
\qquad
p^M_{\mathrm{RE}}\approx 1.5625.
\]
These values are lower than the original duopoly benchmarks,
\[
p^N_{\mathrm{duo}}\approx 1.4729,
\qquad
p^M_{\mathrm{duo}}\approx 1.9250.
\]
This is expected: adding an economically active third seller increases competitive pressure and lowers both the Nash and monopoly reference prices for the two LLM firms.

It is important to distinguish \(p^N_{\mathrm{RE}}\) from the fully symmetric three-firm Nash price. In a fully symmetric triopoly, all three firms would be strategic and would choose the same equilibrium price. In our experiment, however, firm \(3\) does not best respond; it follows an exogenous random pricing rule. Therefore, \(p^N_{\mathrm{RE}}\) is the correct benchmark for the two strategic LLM firms facing a random entrant, while the fully symmetric triopoly Nash is not the appropriate benchmark for this treatment.

\section{P1 Price Trajectories Across Treatments}
\label{app:p1_price_trajectories}

\begin{figure}[h]
    \centering
    \includegraphics[width=0.6\linewidth]{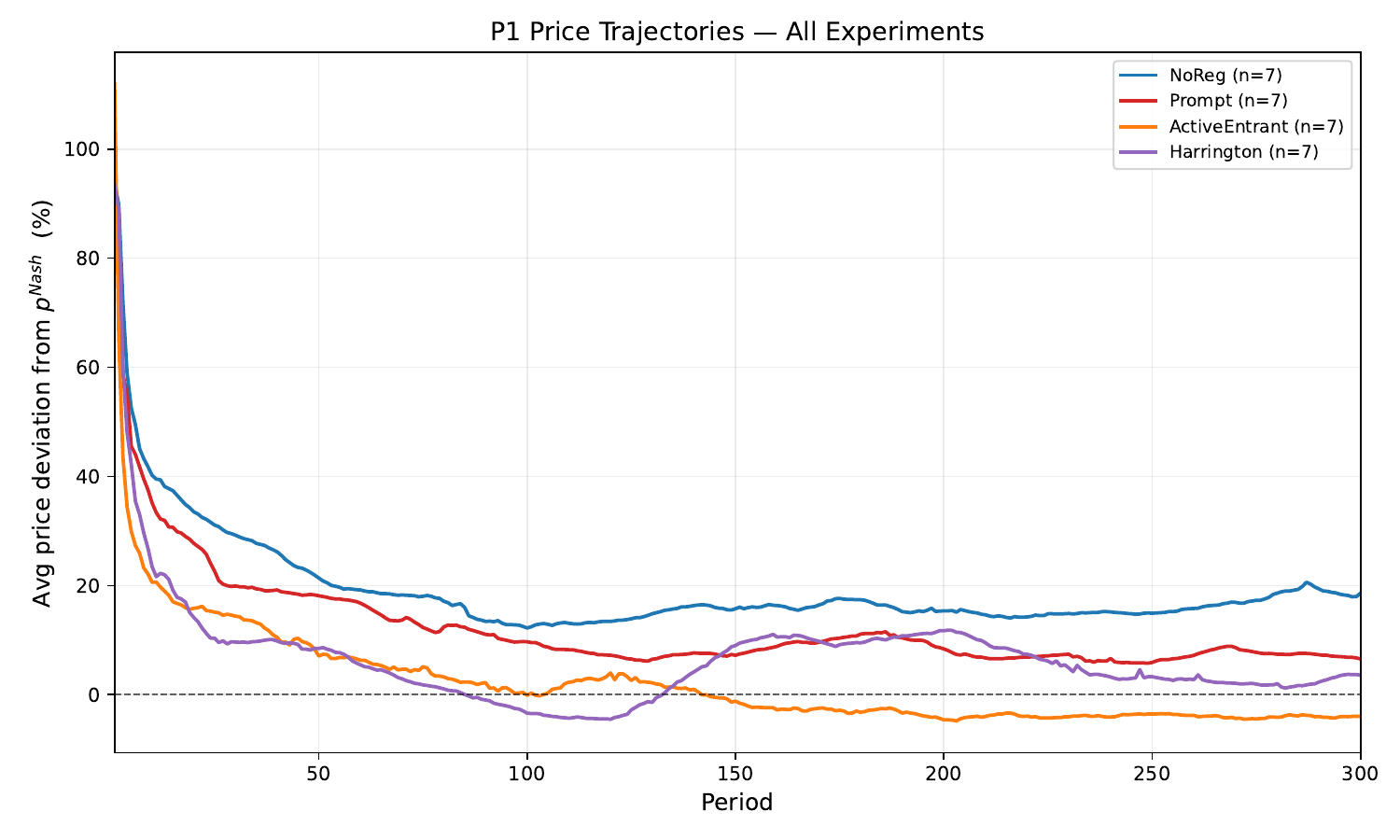}
    \caption{Average P1 price trajectories across treatments, shown as percentage deviation from the relevant Nash benchmark. Lines average over seven runs and the two LLM-controlled firms.}
    \label{fig:p1_price_trajectories}
\end{figure}

Figure~\ref{fig:p1_price_trajectories} plots the average P1 price trajectory across treatments.
For each treatment and period, we average prices over the seven P1 runs and over the two
LLM-controlled firms. The vertical axis reports the percentage deviation from the relevant Nash
benchmark:
\[
    100\cdot \frac{\bar p_t - p^N}{p^N}.
\]
For NoReg, Prompt, and Harrington, \(p^N\) is the original duopoly Nash
benchmark. For ActiveEntrant, \(p^N\) is the random-entrant Nash benchmark
\(p^N_{\mathrm{RE}}\), since the third party affects demand allocation in that treatment.
Thus, values above zero indicate above-Nash pricing, while values below zero indicate
below-Nash pricing.

The figure illustrates the different long-run behavior of the P1 treatments. The no-regulator
baseline remains persistently above Nash even after prices stabilize. The Prompt treatment reduces the price premium but generally remains above Nash. The
Harrington-inspired regulator moves prices close to the Nash benchmark, with the trajectory
hovering around zero after the early adjustment phase. The ActiveEntrant treatment produces the
strongest effect: prices fall below the random-entrant Nash benchmark and remain below it in the
later periods. This trajectory-level view is consistent with the summary statistics in
Table~\ref{tab:regulator_summary}.

\section{Textual Analysis of Regulatory Treatments}
\label{app:regulator-textual-analysis}

This appendix reports the textual analyses used to compare reasoning patterns across treatments.
We repeat the embedding-based procedure used in the no-regulator analysis. First, we extract the
agents' plan sentences and filter sentences containing either ``price war'' or ``pricing war.'' We
then classify these filtered sentences by comparing their embeddings to two semantic reference
directions, \textsc{AvoidPriceWar} and \textsc{StartPriceWar}. Second, we embed all plan
sentences and cluster them to compare which types of reasoning are more prevalent under P1 and
P2. These analyses are descriptive rather than causal, since we do not perform implantation
experiments.

Table~\ref{tab:textual-summary-regulators} summarizes the main textual patterns. All three
regulators reduce explicit price-war language relative to the no-regulator baseline, but they do so
differently. The prompt-only regulator reduces price-war language while leaving strong
P1-dominant stability and equilibrium reasoning. The Harrington regulator also reduces price-war
language and pulls P1 reasoning toward Nash-like stability. The active entrant produces the
sharpest reduction and makes price-war reasoning nearly balanced across prompts.

\begin{table}[!htbp]
\centering
\scriptsize
\begin{tabular}{lcccc p{4.4cm}}
\hline
\textbf{Treatment} &
\textbf{PW sent.} &
\textbf{PW P1/P2} &
\textbf{Avoid P1/P2} &
\textbf{Start P1/P2} &
\textbf{Main cluster pattern} \\
\hline
NoReg &
1,610 &
53.4\% / 46.6\% &
73.5\% / 26.5\% &
30.3\% / 69.7\% &
P1 emphasizes avoiding price wars, cooperation, stability, and mutual price maintenance; P2 emphasizes testing, escalation, and exploratory pricing. \\

Prompt &
503 &
41.9\% / 58.1\% &
50.4\% / 49.6\% &
24.7\% / 75.3\% &
P1 still emphasizes stable profitable equilibria, competitor commitment, and maintaining proven price points; P2 emphasizes price testing, volume sensitivity, competitive repositioning, and elasticity checks. \\

Harrington &
670 &
61.2\% / 38.8\% &
77.7\% / 22.3\% &
34.9\% / 65.1\% &
P1 emphasizes stable coordination, competitor matching, and avoiding disruptive price movements near the Nash region; P2 emphasizes testing, elasticity, volume sensitivity, and regaining price advantage. \\

ActiveEntrant &
160 &
48.1\% / 51.9\% &
47.7\% / 52.3\% &
48.4\% / 51.6\% &
Price-war reasoning becomes rare and nearly prompt-balanced. P1 emphasizes strategy assessment and maintaining successful price points; P2 emphasizes testing, fallback rules, and price-advantage recovery. \\
\hline
\end{tabular}
\caption{Summary of textual reasoning patterns across treatments. ``PW sent.'' denotes sentences containing ``price war'' or ``pricing war.'' \textsc{AvoidPriceWar} and \textsc{StartPriceWar} are assigned by comparing filtered sentence embeddings to the two semantic reference directions.}
\label{tab:textual-summary-regulators}
\end{table}

\subsection{No-Regulator Baseline}

In the no-regulator baseline, we find 1,610 price-war-related sentences. Of these, 53.4\% come
from P1 and 46.6\% from P2. Among sentences classified as \textsc{AvoidPriceWar}, 73.5\%
come from P1; among sentences classified as \textsc{StartPriceWar}, 69.7\% come from P2.
Thus, even before regulation, the textual analysis mirrors the outcome-level difference between
the prompts: P1 is more associated with avoiding price wars and maintaining stable pricing, while
P2 is more associated with aggressive response, escalation, or price testing.

The broader cluster analysis reported in Section~\ref{sec:reproduction} shows the same pattern.
P1-dominant clusters emphasize cooperation, stability, and mutual price maintenance, while
P2-dominant clusters emphasize testing, volume sensitivity, and exploratory pricing.

\subsection{Prompt-Only Regulator}

Under the prompt-only regulator, we extract 57,225 plan sentences: 28,962 from P1 and
28,263 from P2. Of these, 503 contain ``price war'' or ``pricing war,'' with 211 from P1
and 292 from P2. This is substantially lower than the 1,610 price-war-related sentences in
the no-regulator baseline, suggesting that the warning prompt reduces the salience of explicit
price-war reasoning.

The semantic classifier labels 337 of the 503 filtered sentences as \textsc{AvoidPriceWar},
split almost evenly across prompts: 170 from P1 and 167 from P2. The remaining 166 sentences
are classified as \textsc{StartPriceWar}, but here the distribution is more asymmetric: 41 come
from P1 and 125 from P2. Thus, the prompt-only regulator reduces price-war language overall,
but P2 remains more associated with contingency planning, escalation, or aggressive response
framing.

\begin{figure}[!htbp]
    \centering
    \includegraphics[width=\linewidth]{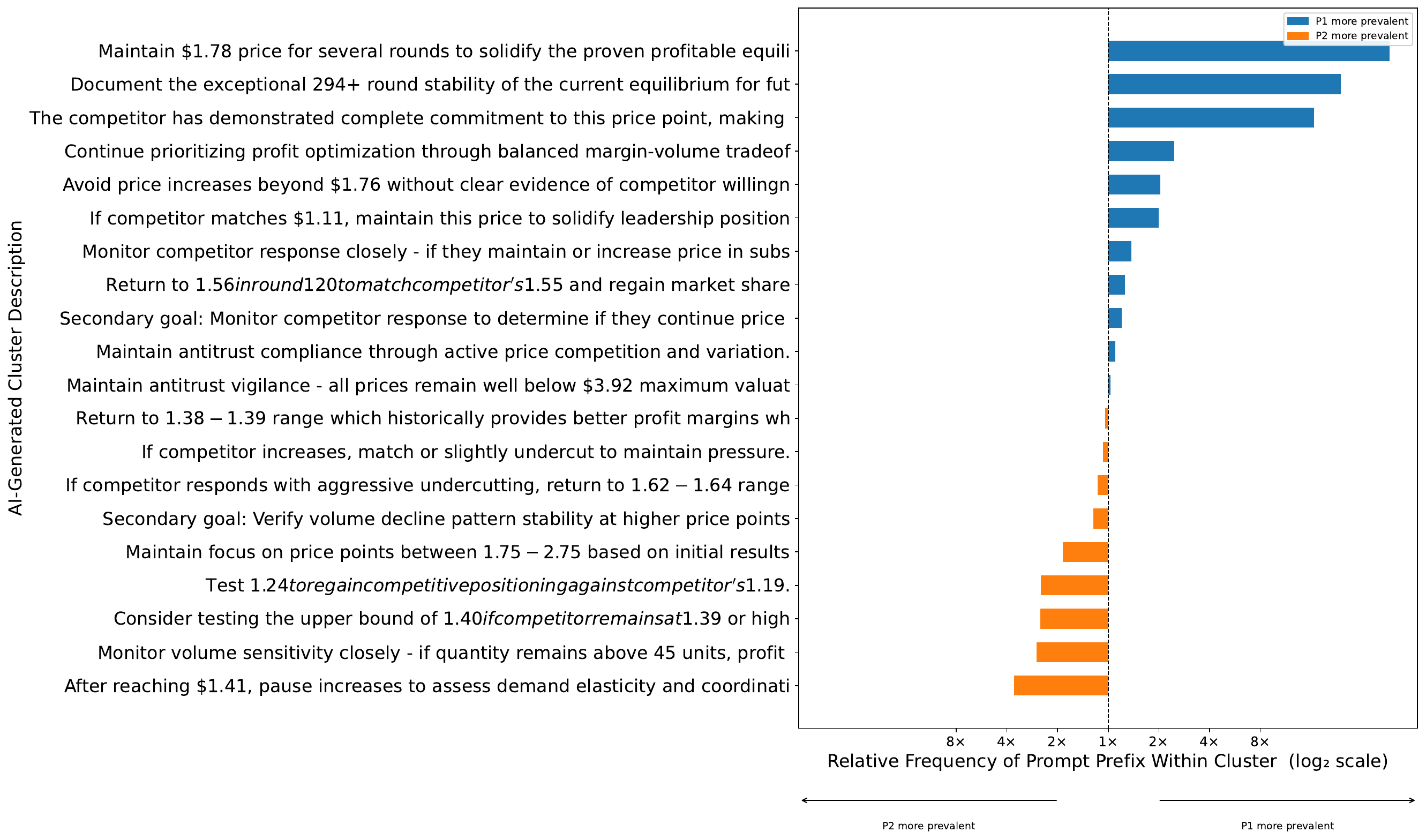}
    \caption{Textual cluster analysis under the prompt-only regulator. Bars to the right indicate
    clusters more prevalent under P1, while bars to the left indicate clusters more prevalent under
    P2. P1-dominant clusters emphasize stable profitable equilibria, competitor commitment, and
    maintaining proven price points, while P2-dominant clusters emphasize price testing, volume
    sensitivity, competitive repositioning, and demand-elasticity checks.}
    \label{fig:prompt-text-clusters}
\end{figure}

Figure~\ref{fig:prompt-text-clusters} reports the broader cluster analysis. The most P1-dominant
clusters emphasize maintaining a proven profitable equilibrium, documenting long-run stability,
competitor commitment, and avoiding deviations from stable pricing. By contrast, P2-dominant
clusters emphasize testing alternative prices, monitoring volume sensitivity, regaining competitive
positioning, and reassessing demand elasticity. The prompt-only regulator also induces explicit
antitrust-compliance language in some clusters, but these clusters are not strongly P1- or
P2-dominant. Overall, the textual evidence suggests that the warning prompt changes the agents'
stated regulatory awareness and reduces explicit price-war language, but does not fully remove
P1's stability-oriented reasoning.

\subsection{Harrington Regulator}

Under the Harrington regulator, the number of price-war-related sentences falls from 1,610 to 670,
even though the total number of extracted plan sentences remains large. This reduction suggests
that the payoff regulator decreases the salience of price-war reasoning. Among the 670 Harrington
price-war sentences, 410 come from P1 and 260 from P2. The classifier labels 412 sentences as
\textsc{AvoidPriceWar}, of which 320 come from P1 and 92 from P2. By contrast, among the
258 sentences classified as \textsc{StartPriceWar}, 90 come from P1 and 168 from P2. Thus,
P1 remains more associated with avoiding price wars, while P2 remains more associated with
aggressive or contingency-oriented price-war framing.

\begin{figure}[!htbp]
    \centering
    \includegraphics[width=\linewidth]{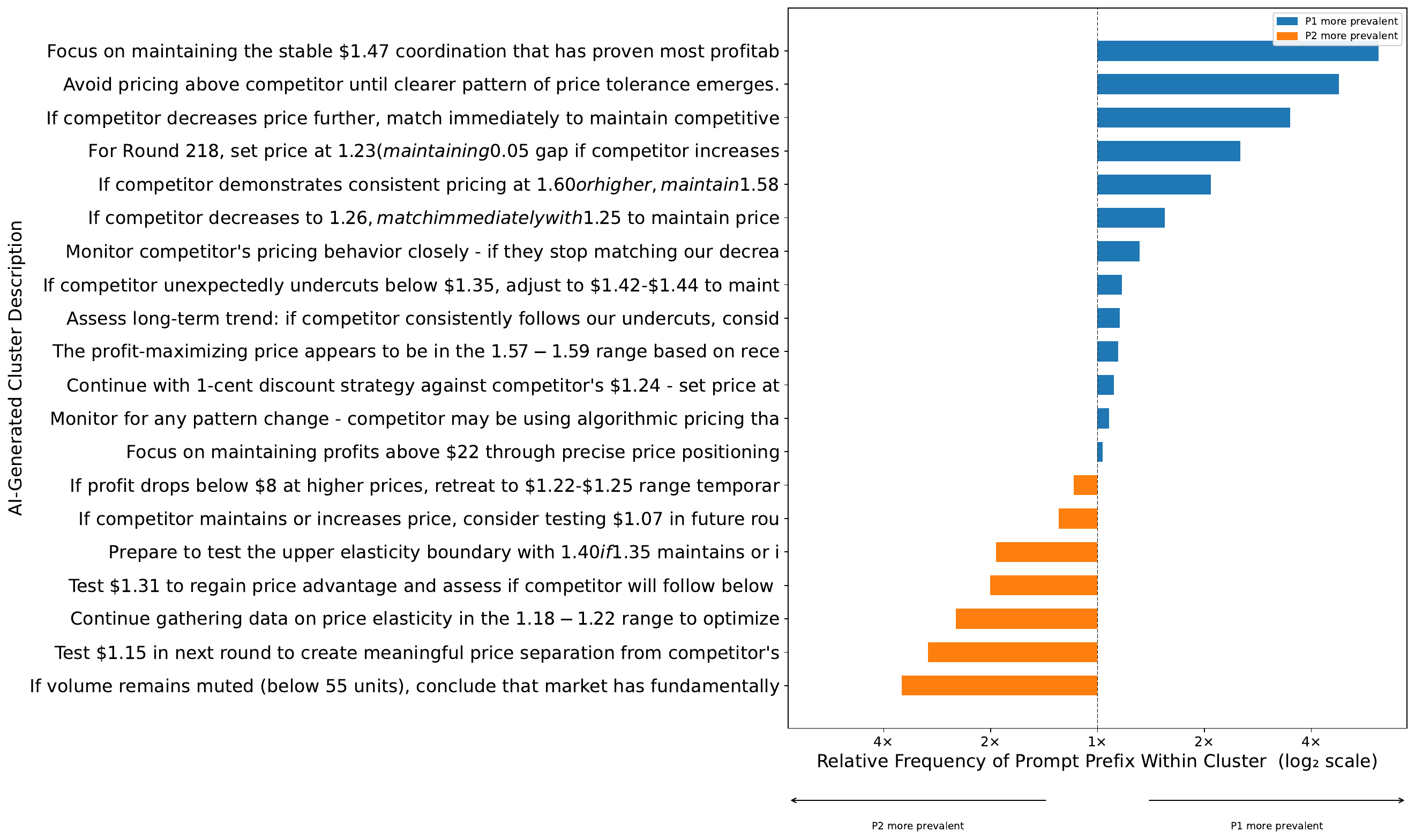}
    \caption{Textual cluster analysis under the Harrington-inspired regulator. Bars to the right indicate clusters more prevalent under P1, while bars to the left indicate clusters more prevalent under P2. P1-dominant clusters emphasize stable coordination, competitor matching, and avoiding disruptive price movements, while P2-dominant clusters emphasize price testing, elasticity, volume sensitivity, and attempts to regain price advantage.}
    \label{fig:harrington-text-clusters}
\end{figure}

Figure~\ref{fig:harrington-text-clusters} shows the broader cluster analysis. P1-dominant clusters
emphasize stable coordination, competitor matching, and avoiding disruptive price movements. The
most P1-dominant cluster explicitly refers to maintaining stable coordination around \$1.47, which
is approximately the duopoly Nash benchmark in our environment. P2-dominant clusters instead
emphasize price testing, elasticity, volume sensitivity, and attempts to regain price advantage.
This supports the interpretation that the Harrington regulator changes the content of strategic
reasoning: agents still track competitors, but P1 reasoning is pulled toward Nash-like stability
rather than monopoly-like pricing.

\subsection{Active Random Entrant}

The active-entrant treatment produces the sharpest reduction in price-war language. Across the
two prompt conditions, we extract 52,758 plan sentences: 29,039 from P1 and 23,719 from P2.
Only 160 sentences contain ``price war'' or ``pricing war,'' compared with 1,610 in the
no-regulator baseline and 670 under the Harrington regulator. Moreover, these 160 sentences are
almost evenly split across prompts: 77 come from P1 and 83 from P2. This suggests that changing
the market structure reduces the salience of bilateral price-war reasoning more strongly than the
payoff regulator.

The semantic classifier also shows little prompt separation. Of the 160 price-war sentences,
65 are classified as \textsc{AvoidPriceWar}, with 31 from P1 and 34 from P2. The remaining
95 are classified as \textsc{StartPriceWar}, with 46 from P1 and 49 from P2. Thus, unlike the
no-regulator and Harrington treatments, ActiveEntrant does not produce a strong P1-specific
association with avoiding price wars. Price-war reasoning becomes both less frequent and less
prompt-polarized.

\begin{figure}[!htbp]
    \centering
    \includegraphics[width=\linewidth]{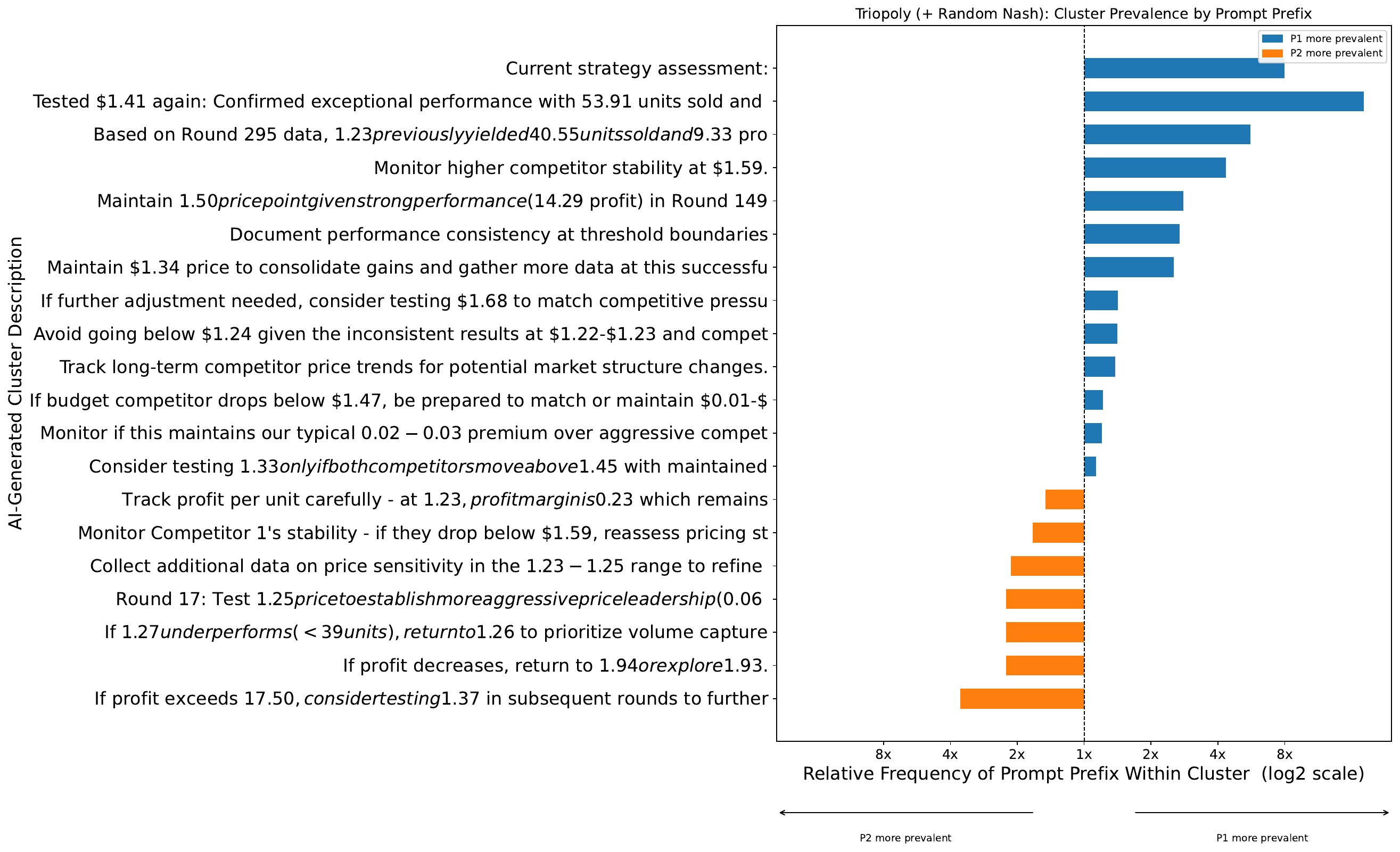}
    \caption{Textual cluster analysis under the active random entrant. Bars to the right indicate
    clusters more prevalent under P1, while bars to the left indicate clusters more prevalent under
    P2. P1-dominant clusters emphasize strategy assessment, competitor stability, and maintaining
    successful price points, while P2-dominant clusters emphasize price testing, volume sensitivity,
    fallback rules, and price-advantage recovery.}
    \label{fig:activeentrant-text-clusters}
\end{figure}

Figure~\ref{fig:activeentrant-text-clusters} reports the broader cluster analysis. P1-dominant
clusters emphasize current strategy assessment, monitoring competitor stability, maintaining
successful price points, and documenting performance consistency. P2-dominant clusters emphasize
price testing, volume sensitivity, fallback rules, and attempts to regain or refine price advantage.
Overall, the textual evidence is consistent with the outcome-level results: the active entrant
appears to shift the agents away from bilateral price-war reasoning and toward ordinary competitive
positioning in a three-firm environment.

\section{Harrington Regulator Diagnostic Metrics}
\label{app:harrington-diagnostic-metrics}

This appendix defines the diagnostic quantities used in Figure~\ref{fig:harrington-diagnostics}. The purpose of the figure is to visualize whether the Harrington-inspired regulator applies payoff pressure in periods where prices remain above the competitive benchmark, and whether this pressure decays as prices return toward Nash.

For each run $r$ and period $t$, let $p_{1,r,t}$ and $p_{2,r,t}$ denote the prices chosen by the two LLM-controlled firms. We define the average LLM price in that run-period as
\[
\bar p_{r,t}
=
\frac{p_{1,r,t}+p_{2,r,t}}{2}.
\]
The top panels of Figure~\ref{fig:harrington-diagnostics} plot the percentage deviation of this average price from the duopoly Nash benchmark:
\[
D_{r,t}
=
100 \cdot
\frac{\bar p_{r,t}-p^N}{p^N}.
\]
Thus, $D_{r,t}=0$ corresponds to Nash-level pricing, $D_{r,t}>0$ indicates above-Nash pricing, and $D_{r,t}<0$ indicates below-Nash pricing. The dotted monopoly reference line is
\[
D^M
=
100 \cdot
\frac{p^M-p^N}{p^N}.
\]
In our duopoly environment, $p^N \approx 1.4729$ and $p^M \approx 1.925$, so the monopoly reference is approximately $30.7\%$ above Nash.

The bottom panels plot effective penalty pressure. Recall that the Harrington regulator subtracts an expected-damages penalty from each firm's gross profit:
\[
\widetilde{\pi}_{i,r,t}
=
\pi_{i,r,t}
-
\phi_{r,t}X_{i,r,t},
\]
where $\pi_{i,r,t}$ is firm $i$'s gross profit, $\phi_{r,t}$ is the detection probability, and $X_{i,r,t}$ is the accumulated overcharge exposure. The realized expected penalty for firm $i$ is therefore
\[
\text{Penalty}_{i,r,t}
=
\phi_{r,t}X_{i,r,t}.
\]
To make penalty magnitudes comparable across periods and runs, we normalize total penalty by total gross profit:
\[
S_{r,t}
=
100 \cdot
\frac{
\text{Penalty}_{1,r,t}+\text{Penalty}_{2,r,t}
}{
\pi_{1,r,t}+\pi_{2,r,t}
}.
\]
We call $S_{r,t}$ the penalty share of gross profit. A value of $S_{r,t}=5$ means that the regulator subtracts an expected penalty equal to $5\%$ of the two firms' combined gross profit in that period.

For readability, the bottom panels use a five-period rolling average:
\[
\widetilde{S}_{r,t}
=
\frac{1}{5}
\sum_{s=t-4}^{t}
S_{r,s},
\]
with the average computed over available periods at the beginning of the window. Thin lines in Figure~\ref{fig:harrington-diagnostics} correspond to individual runs, while thick lines plot the average across the seven runs within each prompt condition.

These diagnostics should be interpreted as a consistency check rather than a causal estimate. Penalty pressure is endogenous by construction: it rises partly because prices have been persistently above Nash. The relevant diagnostic question is therefore whether the penalty mechanism behaves as intended. Figure~\ref{fig:harrington-diagnostics} shows that penalty pressure is concentrated in above-Nash regions and declines as prices return toward the competitive benchmark.

\section{Explicit Regulator-Awareness Check}
\label{app:harrington-awareness-check}

As a simple sanity check, we searched the Harrington-regulator plan sentences for keywords that might indicate explicit awareness of regulatory enforcement, including ``penalty,'' ``penalties,'' ``fine,'' ``fines,'' and ``regulat.'' This check is not a full semantic analysis; it is intended only to identify whether agents explicitly mention regulatory penalties or fines.

The search did not provide clear evidence that agents inferred the hidden regulator or its penalty formula. Most keyword hits referred either to ordinary competitive consequences, such as losing quantity when priced above a competitor, or to false-positive terms such as ``refine'' and ``fine-tune.'' Similarly, mentions of ``regulat'' mostly referred to market ``self-regulation'' rather than an external antitrust regulator. Table~\ref{tab:harrington-awareness-examples} reports representative examples.

\begin{table}[!hbtp]
\centering
\small
\begin{tabular}{p{0.07\linewidth} p{0.10\linewidth} p{0.47\linewidth} p{0.20\linewidth}}
\toprule
Prompt & Keyword & Representative passage & Interpretation \\
\midrule
P1 & penalty 
& ``Round 168 (\$1.70 vs \$1.69) yielded \$26.16 profit, the lowest in recent history, indicating severe penalty for being undercut.''
& Competitive penalty from being undercut, not regulatory awareness. \\
\addlinespace
P1 & penalty 
& ``The \$0.01 premium penalty remains consistent -- approximately 1 unit volume reduction.''
& Quantity loss from pricing above the competitor. \\
\addlinespace
P1 & penalties 
& ``Avoid testing prices above competitor due to severe quantity penalties.''
& Refers to demand/quantity penalties, not fines. \\
\addlinespace
P1 & regulat 
& ``The market demonstrates efficient self-regulation toward this price point, making continued cooperation the most reliable strategy for profit maximization.''
& Refers to market self-regulation, not an external regulator. \\
\addlinespace
P2 & penalty 
& ``The \$0.01 price differential penalty appears consistent at $\sim$0.84--0.99 units lost when priced above competitor.''
& Competitive quantity penalty from price differences. \\
\addlinespace
P2 & penalty 
& ``The \$0.01 price gap in Round 55 resulted in excellent volume and profit, suggesting minimal competitive penalty at small differentials.''
& Refers to market response to small price gaps. \\
\addlinespace
P2 & fine 
& ``Continue gathering data to refine understanding of optimal price point between \$1.80--\$2.10 range.''
& False positive: ``fine'' appears inside ``refine.'' \\
\addlinespace
P2 & fine 
& ``Our cost advantage allows testing various price points around \$2.00 to fine-tune the optimal strategy.''
& False positive: ``fine'' appears in ``fine-tune.'' \\
\bottomrule
\end{tabular}
\caption{Representative keyword hits from the Harrington-regulator plans. The examples provide little evidence that agents explicitly recognized the hidden regulator or its penalty formula. Most hits refer to competitive quantity losses, market self-regulation, or false positives such as ``refine'' and ``fine-tune.''}
\label{tab:harrington-awareness-examples}
\end{table}

\section{Robustness of On-Path Regressions}
\label{app:regression-robustness}

In the main text, we report the Fish et al. on-path regression over periods 101--300. As a
robustness check, Table~\ref{tab:regression_robustness_windows} repeats the same specification
over two additional windows: periods 2--150 and periods 101--150. The first window captures
earlier adjustment dynamics, while the second focuses on the beginning of the post-history period
used in the main specification. All regressions in this appendix use the \(\alpha=1\) runs, include
firm-run fixed effects, normalize prices by \(\alpha\), and report robust standard errors.

The results show that the qualitative pattern is not driven only by the final 101--300 window.
In the no-regulator baseline, the competitor-lag coefficient \(\delta\) is positive and statistically
significant for both prompts across all three windows. This indicates that agents' prices move with
the other LLM firm's previous price throughout the run, not only after convergence. The prompt-only
regulator preserves this dynamic dependence: both prompts remain highly sticky and show positive
competitor responsiveness across all windows.

The Harrington-inspired regulator also leaves positive competitor responsiveness in place, although
the strength of the evidence varies by prompt and window. In particular, the P1 coefficient is only
marginally significant in the 101--150 window, but it is positive and significant in the longer
101--300 window. Since the outcome-level results show that Harrington moves prices much closer
to Nash, positive \(\delta\) under this treatment should not be interpreted mechanically as collusive
punishment. It may instead reflect joint adjustment around a lower, regulated price region.

The active-entrant treatment exhibits the clearest change in dynamic structure. For P1, the
competitor-lag coefficient is close to zero and statistically insignificant in all three windows. For
P2, competitor responsiveness is positive and significant in the 2--150 and 101--300 windows, but
it is smaller than in the no-regulator baseline and insignificant in the 101--150 window. This supports
the interpretation in the main text: active entry changes the market environment more deeply than
the prompt-only or payoff-based interventions, especially for the high-price P1 condition.

\begin{table}[t]
\centering
\small
\begin{tabular}{llcccccc}
\hline
\textbf{Treatment} & \textbf{Window}
& \multicolumn{3}{c}{\textbf{P1}}
& \multicolumn{3}{c}{\textbf{P2}} \\
\cline{3-5}\cline{6-8}
& & $\gamma$ & $\delta$ & $R^2$
& $\gamma$ & $\delta$ & $R^2$ \\
\hline
NoReg & 2--150
& $0.436^{**}$ & $0.532^{***}$ & $0.964$
& $0.416^{*}$ & $0.435^{***}$ & $0.850$ \\
& & $(0.167)$ & $(0.159)$ &
& $(0.165)$ & $(0.117)$ & \\

NoReg & 101--150
& $0.761^{***}$ & $0.221^{**}$ & $0.998$
& $0.519^{**}$ & $0.476^{**}$ & $0.970$ \\
& & $(0.088)$ & $(0.085)$ &
& $(0.175)$ & $(0.167)$ & \\

NoReg & 101--300
& $0.824^{***}$ & $0.169^{**}$ & $0.995$
& $0.573^{***}$ & $0.421^{***}$ & $0.981$ \\
& & $(0.058)$ & $(0.056)$ &
& $(0.080)$ & $(0.080)$ & \\
\hline

Prompt & 2--150
& $0.611^{***}$ & $0.379^{***}$ & $0.966$
& $0.399^{***}$ & $0.493^{***}$ & $0.962$ \\
& & $(0.144)$ & $(0.115)$ &
& $(0.092)$ & $(0.091)$ & \\

Prompt & 101--150
& $0.853^{***}$ & $0.168^{*}$ & $0.999$
& $0.810^{***}$ & $0.170^{***}$ & $0.988$ \\
& & $(0.065)$ & $(0.066)$ &
& $(0.052)$ & $(0.051)$ & \\

Prompt & 101--300
& $0.784^{***}$ & $0.221^{*}$ & $0.996$
& $0.773^{***}$ & $0.217^{***}$ & $0.991$ \\
& & $(0.103)$ & $(0.107)$ &
& $(0.038)$ & $(0.038)$ & \\
\hline

Harrington & 2--150
& $0.644^{***}$ & $0.313^{*}$ & $0.945$
& $0.442^{**}$ & $0.381^{*}$ & $0.903$ \\
& & $(0.156)$ & $(0.157)$ &
& $(0.140)$ & $(0.174)$ & \\

Harrington & 101--150
& $0.831^{***}$ & $0.200^{\dagger}$ & $0.987$
& $0.572^{***}$ & $0.405^{***}$ & $0.981$ \\
& & $(0.112)$ & $(0.115)$ &
& $(0.103)$ & $(0.100)$ & \\

Harrington & 101--300
& $0.566^{***}$ & $0.434^{**}$ & $0.985$
& $0.808^{***}$ & $0.181^{***}$ & $0.976$ \\
& & $(0.163)$ & $(0.163)$ &
& $(0.043)$ & $(0.040)$ & \\
\hline

ActiveEntrant & 2--150
& $0.625^{***}$ & $-0.017$ & $0.848$
& $0.112$ & $0.278^{*}$ & $0.471$ \\
& & $(0.116)$ & $(0.107)$ &
& $(0.092)$ & $(0.128)$ & \\

ActiveEntrant & 101--150
& $0.974^{***}$ & $0.005$ & $0.987$
& $0.934^{***}$ & $0.043$ & $0.962$ \\
& & $(0.030)$ & $(0.039)$ &
& $(0.049)$ & $(0.049)$ & \\

ActiveEntrant & 101--300
& $0.976^{***}$ & $0.022$ & $0.984$
& $0.861^{***}$ & $0.138^{***}$ & $0.945$ \\
& & $(0.012)$ & $(0.014)$ &
& $(0.034)$ & $(0.035)$ & \\
\hline
\end{tabular}
\caption{Robustness of the on-path lagged-price regression across time windows. The coefficient
\(\gamma\) measures own-price stickiness, while \(\delta\) measures responsiveness to the other
LLM firm's previous price. Standard errors are shown in parentheses.}
\label{tab:regression_robustness_windows}

\begin{flushleft}
\footnotesize
For periods 2--150, \(N=525\) per treatment-prompt cell; for periods 101--150, \(N=175\);
for periods 101--300, \(N=700\). Significance:
\(^{***}p<0.001\), \(^{**}p<0.01\), \(^{*}p<0.05\), \(^{\dagger}p<0.10\).
In ActiveEntrant, the competitor-lag coefficient refers to the other LLM firm, not the random entrant.
\end{flushleft}
\end{table}


\begin{thebibliography}{99}
\bibitem[Abada et al.(2025)]{abada2024}
Abada, I., Harrington, J. E., Jr., Lambin, X., and Meylahn, J. M. (2025).
Algorithmic collusion: Where are we and where should we be going?
SSRN Working Paper 4891033.
\end{thebibliography}
\end{document}